\documentclass[twocolumn, pre]{revtex4-1}
\usepackage[toc,page]{appendix}
\usepackage{color}
\definecolor{red}{rgb}{0.75,0,0}
\definecolor{blue}{rgb}{0,0,0.75}
\definecolor{green}{rgb}{0,0.5,0}
\definecolor{yellow}{rgb}{0,0.5,0}
\usepackage[pdftex, pdfborder={0 0 0}, colorlinks=true, linkcolor=red, urlcolor=blue]{hyperref}
\usepackage{mathrsfs}
\usepackage{notes2bib}
\usepackage{graphicx,latexsym,setspace,lipsum}
\usepackage{amsmath}
\usepackage{amssymb}
\usepackage{bm}
\usepackage{wrapfig}
\usepackage{graphicx}
\usepackage{graphicx, caption, subcaption}

\usepackage[usenames,dvipsnames,svgnames,table]{xcolor}

\def\be{\begin{equation}}
\def\ee{\end{equation}}
\def\bea{\begin{eqnarray}}

\def\a{\alpha}
\def\b{\beta}

\def\besub{\begin{subequations}}
\def\eesub{\end{subequations}}

\def\P{{\bf P}}

\def\bwd{\begin{widetext}}
\def\ewd{\end{widetext}}

\newcommand{\bigzero}{\mbox{\normalfont\bfseries 0}}
\newcommand{\bigj}{\mbox{\normalfont\bfseries J}}
\newcommand{\rvline}{\hspace*{-\arraycolsep}\vline\hspace*{-\arraycolsep}}
\begin{document}
%\title{{Aligned active suspensions: superdiffusion and anisotropic criticality}}	
\title{{Time-reversal symmetries and equilibrium-like Langevin equations}}
\author{Lokrshi Prawar Dadhichi and Klaus Kroy}
\email{lpdadhichi@gmail.com}
\email{klaus.kroy@uni-leipzig.de}
\affiliation{Institute for Theoretical Physics, Leipzig University, 04103 Leipzig, Germany}

\begin{abstract}
Graham  has shown in Z. Physik B 26, 397-405 (1977)
% and later inspired by his work Eyink et al., J Stat Phys  385-472 (1996), 
that a fluctuation-dissipation relation can be imposed on a class of non-equilibrium Markovian Langevin equations that admit a stationary solution of the corresponding Fokker-Planck equation.
% be written in "Onsager force-flux form", where force in turn can be written as derivative of a potential. 
The resulting equilibrium form of the Langevin equation is associated with a nonequilibrium Hamiltonian. {Here we provide some explicit insight into how this Hamiltonian} {may loose its time reversal invariance and how} the ``reactive'' and ``dissipative'' fluxes loose their distinct time reversal symmetries. The antisymmetric coupling matrix between forces and fluxes no longer originates from Poisson brackets and the ``reactive'' fluxes contribute to the (``housekeeping'') entropy production, in the steady state. The time-reversal even and odd parts of the nonequilibrium Hamiltonian contribute in qualitatively different but physically instructive ways to the entropy. {We find instances where fluctuations due to noise are solely responsible for the dissipation.} Finally, this structure gives rise to a new, physically pertinent instance of frenesy.
\end{abstract}

\maketitle

\section{Introduction}
\label{intro}
A Langevin equation is a stochastic differential equation describing generic mesoscopic dynamics driven by {both} systematic and stochastically fluctuating forces \cite{zwanzig2001nonequilibrium}. The concept is suitable for slow degrees of freedom coupled { {to a}} large number of fast degrees of freedom that can be subsumed into {a} ``noisy'' stochastic force. There are multiple ways to derive Langevin equations from microscopic descriptions \cite{zwanzig2001nonequilibrium,
mori1965transport}. In equilibrium, the deterministic part of the dynamics is governed by an effective (i.e., typically coarse grained) Hamiltonian. It therefore relies on a crucial feature of equilibrium dynamics, namely that the coarse-graining, by which one exploits the scale separation between the slow systematic and fast stochastic degrees of freedom, leads one to a free energy governing the slow variables that itself obeys Hamiltonian symmetries.  In this case, the stochastic noise strength can moreover be fully specified mesoscopically, by a  fluctuation-dissipation relation (FDR) \cite{kubo1966fluctuation,kubo2012statistical,balakrishnan2008elements} that obliges the fast degrees of freedom to act as an effective thermostat for the slow variables, so that the solutions obtained for the latter from the Langevin equation coincide with {those from} the classical Gibbs ensembles, at late times. 
In this framework, dissipative and reactive (reversible/conservative) contributions can be clearly distinguished. 

Traditionally, the FDR is thus intimately associated with thermal equilibrium \cite{kubo1966fluctuation,kubo2012statistical,balakrishnan2008elements} and its failure with a loss of equilibrium. Indeed, a set of Langevin equations describing a generic {nonequilibrium} system is not obliged to obey Hamiltonian dynamics nor any FDR. However, the FDR has time and again been generalized {to nonequilibrium conditions} ---\emph{{pars pro toto}} we here refer to {Refs.}~\cite{hanggi1978stochastic,
falcioni1990correlation,cugliandolo1994off,ruelle1998general,
nakamura2008fluctuation,speck2006restoring,
prost2009generalized,lippiello2005off,
sarracino2019fluctuation,dal2021fluctuation}, and references therein. Interestingly Graham \cite{graham1977covariant} and {later,} inspired by his work, Eyink et al. \cite{eyink1996hydrodynamics} gave a formal procedure to extend the FDR to generic nonequilibrium mesoscopic Markov systems, whenever the existence of a stationary solution of the associated Fokker--Planck equation (FPE) can be taken for granted. In principle, it provides a formal effective Hamiltonian-like structure reminiscent of a potential of mean force \cite{hansen2013theory}, given by the logarithm of  said stationary solution of the FPE, to which we want to refer as the nonequilibrium Hamiltonian (NH). As we recall and explicitly lay out in the following, the nonequilibrium Langevin equations rephrased in terms of this NH have the same structure as in equilibrium. {The NH is a Lyponov function for the Langevin dynamics around the steady state, so that the latter is unique and stable \cite{graham1977covariant}.} The resemblance of the equations with the equilibrium structure, including a formal FDR, naturally raises the question, where the condition of nonequilibrium got hidden? {As pointed out in Refs.~\cite{graham1977covariant,eyink1996hydrodynamics}, it is hidden in the symmetry under time reversal of the dynamical equation. Here, we dwell deeper into this question and, through an exactly solvable model (studied widely in active matter), demonstrate how ``dissipative" and ``reactive" parts of the nonequilibrium Langevin equations violate the familiar equilibrium time-reversal signatures.}  {Additionally we show that the  time-reversal  even and odd part of the NH contribute to the entropy production in qualitatively different ways. Apart from being a Lyponov function for a given steady state, the (negative) NH is also related to the entropy and the excess heat produced in a quasistatic operation turning it into another steady state \cite{hatano2001steady}. This provides multiple reasons to study the effect of perturbing the NH. } Unlike the usual practice for nonequilibrium systems, where perturbing forces are directly added to the equations of motion (which can be understood as a force balance), here the perturbing force appears in the NH with a special coupling \cite{graham1977covariant}. In this context, we establish the {link to  frenesy} \cite{maes2020frenesy,maes2020response}, which is a measure of the ``undirected'' currents in a system. 

More precisely, we show that, in such equilibrium-like Langevin equations, the nonequilibrium condition manifests itself in the following  ways:
\begin{itemize}
  \item{The NH {need not be} time reversal invariant.}
    \item{The antisymmetric couplings do not arise from Poisson brackets.}
    \item{The ``reactive'' currents {also} produce  entropy.} { Similarly,  fluctuations from the steady state, due to the  noise, can produce entropy (``active/dissipative" fluctuations).}
\end{itemize}
We also make the following important observations:
\begin{itemize}
  \item {The parts of the NH with different time-reversal signatures contribute in qualitatively different ways to the entropy production.}
  \item {The NH opens a new meaningful way to perturb the system and hence provides a second interesting instance of excess frenesy, beyond the usual one.} {We also point out that this perturbation can be used to derive a variant of, the Harada-Sasa relation \cite{harada2005equality}.}
\end{itemize}

The paper is organised as follows: in Sec.~\ref{sec1} we recall the structure of equilibrium Langevin equations, Sec.~\ref{sec2} summarizes Graham's work \cite{graham1977covariant} and establishes the structural similarity between Langevin equations with nonequilibrium steady states and equilibrium Langevin equations. In Sec.~\ref{sec3}, we analytically solve the FPE corresponding to a linear nonequilibrium Langevin equation to explicitly reveal this equilibrium-like structure. We discuss its interesting features, and, in Sec.~\ref{sec4}, apply this formalism to a much studied model in the physics of soft active matter, namely so-called active Ornstein--Uhlenbeck particles (AOUPs) \cite{fodor2016far}. {In Sec.~\ref{sec5}, the role of different parts of NH in the entropy production is studied.} \ Finally, Sec.~\ref{sec6} provides the link to frenesy in this context and a comparison with previous studies~\cite{maes2020response}.

\section{Equilibrium structure}
\label{sec1}
 We construct equations of motion for dynamical variables $\bm{\mathcal{C}}$ with position-like and momentum-like components $\bm{\mathcal{Q}}$ and $\bm{\mathcal{P}}$, respectively, even and odd under time-reversal (hereafter denoted by $\mathcal{T}$). We allow
%\cite{kumar2008active}}
$\bm{\mathcal{C}}$ to be finite or infinite-dimensional, depending on whether we are dealing with a system parameterized in terms of particle degrees of freedom or with a spatially extended system described by a stochastic field theory. In particle systems in thermal equilibrium, $\bm{\mathcal{Q}}$ and $\bm{\mathcal{P}}$ normally refer to canonically conjugate variables, but in more strongly coarse-grained formulations \cite{chaikin1995principles}, and in particular in generalized non-equilibrium Langevin systems, they do not have to, nor do they need to have the same number of components. The stochastic equations of motion describing the \emph{thermal equilibrium} dynamics of $\bm{\mathcal{C}}$ are   \cite{ma1975sk,chaikin1995principles,zwanzig2001nonequilibrium,lau2007state}
\begin{equation}
\label{eq:genlang}
\partial_t\bm{\mathcal{C}}=-(\bm{\Gamma}+\bm{\mathcal{W}})\cdot\partial_{\bm{\mathcal{C}}}H+T\partial_{\bm{\mathcal{C}}}\cdot\bm{\mathcal{W}} +\bm{\xi}
\end{equation}
where $H(\mathcal{C})$ is the effective Hamiltonian,  $\bm{\Gamma}$ is a symmetric matrix of dissipative couplings between the variables that governs the FDR
\begin{equation}
\label{eq:gennoise}
\langle\bm{\xi}(t)\bm{\xi}(t'){{\rangle}}=2T\bm{\Gamma}\delta(t-t')\,,
\end{equation}
and $\bm{\mathcal{W}}$ is an antisymmetric matrix of reactive couplings. For brevity, we are using the notation for discrete degrees of freedom, which can however straightforwardly be upgraded for the case that $\bm{\mathcal{C}}$ is supposed to be a field variable; e.g., the  term $\bm{\mathcal{W}}\cdot\partial_{\bm{\mathcal{C}}}H$ would read 
\begin{equation}
 \int d\mathbf{x}'\mathcal{W}_{\mu\nu}(\mathbf{x},\mathbf{x}')\frac{\delta H}{\delta\mathcal{C}_\nu(\mathbf{x'})}
\end{equation}
(summation over $\nu$ implied).

Terms involving $\bm{\mathcal{W}}$ must have,  
 component by component, the same signature under $\mathcal{T}$ as $\partial_t\bm{\mathcal{C}}$, and those involving $\bm{\Gamma}$ must have the opposite $\mathcal{T}$-signature. Thus the $\bm{\mathcal{Q}} \bm{\mathcal{Q}}$ and $\bm{\mathcal{P}} \bm{\mathcal{P}}$ components of {the matrix} $\bm{\Gamma}$ must themselves be even under $\mathcal{T}$, while {its} $\bm{\mathcal{Q}} \bm{\mathcal{P}}$ and $\bm{\mathcal{P}} \bm{\mathcal{Q}}$  components must be odd. In equilibrium, $\bm{\mathcal{W}}$ is identified with the Poisson bracket between the dynamical variables as discussed later in this section \cite{chaikin1995principles,ma1975sk}.

%The case of specific interest to us is where $\bm{\mathcal{Q}}$ consists of a spatial part ${\bf X}$ and the chemical coordinate $n$, and $\bm{\mathcal{P}}$ has only a spatial part ${\bf P}$. Then the components $\bm{\Gamma}_{{\bf X}{\bf X}}$, $\bm{\Gamma}_{{\bf X}n}$, $\bm{\Gamma}_{\bf{P}\bf{P}}$, and $\Gamma_{nn}$ must be even under $\mathcal{T}$, and $\bm{\Gamma}_{{\bf X}{\bf P}}$ and $\bm{\Gamma}_{{\bf P}n}$ must be odd. Since we are not considering the possibility of an external field that breaks {$\mathcal{T}$}, such as a magnetic field \cite{casimir1945onsager,de2013non}, this implies that $\bm{\Gamma}_{{\bf X}{\bf P}}$, and $\bm{\Gamma}_{{\bf P}n}$ should themselves be odd in ${\bf P}$. 
%Since we are not considering the possibility of an external field that breaks $\mathcal{T}$, such as a magnetic field \cite{casimir1945onsager,de2013non}, this implies that $\bm{\Gamma}_{{\bf X}{\bf P}}$, and $\bm{\Gamma}_{{\bf P}n}$ should themselves be odd in $\bf P$.

Standard derivations of generalized Langevin equations \cite{mori1965transport,ma1975sk,chaikin1995principles,mazenko2008nonequilibrium}  require  the additional term $T{\nabla}_{\bm{\mathcal{C}}}\cdot \bm{\mathcal{W}}$ in Eq.~\eqref{eq:genlang} for the steady state to be $\propto$ $e^{-H/T}$. While it  vanishes in familiar equilibrium Langevin equations \cite{hohenberg1977theory}, there are natural instances in active matter where it is nonzero \cite{dadhichi2018origins}. 
 Moreover, when ${\mathbf{\Gamma}}$ depends on $\bm{\mathcal{C}}$, the noise {term in} \eqref{eq:genlang} is multiplicative. 
It then produces a spurious drift. For the steady-state distribution to remain $e^{-H/T}$, we must then include, as a counter term, the additional drift $T (\nabla_{\bm{\mathcal{C}}} \cdot {\bf \Gamma} - \alpha \bf g\cdot \nabla_{\bm{\mathcal{C}}} g)$,  in Eq.~\eqref{eq:genlang}, where $\mathbf{g} \cdot \bf g = 2 {\bf \Gamma}$ \cite{lau2007state}, and the continuous parameter $\alpha\in[0,1]$ parameterizes different physical interpretations of the noise. Similarly, {$\mathcal{W}_{{{\bf\mathcal{P}}}{\mathcal{P}}}$} should be odd under {$\mathcal{T}$} and therefore suitably $\bm{\mathcal P}$-dependent.  

For an equilibrium system, the reactive (reversible) term emerges from the Poisson bracket of the variable with the Hamiltonian \citep{chaikin1995principles, ma1975sk}.
\begin{equation}
\begin{split}
\partial_t{\mathcal{C}}_\mu & =\{H, \mathcal{C}_\mu\}\equiv -\mathcal{W}_{\mu\nu}\partial_{\mathcal{C}_\nu}H
\end{split}
\end{equation}
The antisymmetric coupling matrix $\mathcal{W}_{\mu\nu}=-\mathcal{W}_{\nu\mu}=\{\mathcal{C}_\mu,\mathcal{C}_\nu\}$ has the structure of a Poisson bracket of the dynamical (field) variables.
And, again, an extra term $\partial_\mu\mathcal{W}_{\mu\nu}$ is required in the reactive part to attain a Boltzmann equilibrium distribution. Hydrodynamic Poisson brackets are usually calculated directly from a microscopic model \cite{stark2005poisson} or indirectly inferred from symmetries \cite{dzyaloshinskii1980poisson}.

In general, the above effective Hamiltonian structure, and hence the clear identification of reactive and dissipative currents, breaks down for nonequilibrium systems, which naturally raises the question how much of it can be rescued for the special subclass of Markov systems that admit non-equilibrium steady states (NESS). 

\section{Graham's equilibrium-like structure}
\label{sec2}
Graham \cite{graham1977covariant} and later, inspired by his work, Eyink \emph{et al.~}\citep{eyink1996hydrodynamics} gave a formal procedure how to write nonequilibrium Markovian equations in an equilibrium-like form, in which   Eqs.~(\ref{eq:genlang}), (\ref{eq:gennoise}) still pertain. In particular, Eyink \emph{et al.~}pointed out that the ``dissipative'' (symmetric) coupling governing the strength of the noise correlation actually establishes a FDR 
of the first {kind}, as discussed further below.
Here we summarise the results that are useful for our present purpose. Namely, a general Langevin equation (for discrete degrees of freedom)
\begin{align}
\label{lang}
\dot{\mathcal{C}}_\mu=J_\mu(\bm{\mathcal{C}})+g_\mu^i(\bm{\mathcal{{{C}}}})\xi_i
\end{align}
where
\begin{align}
\label{noi1}
\langle \xi_i(t)\xi_j(t')\rangle=2\delta_{ij}\delta(t-t') \quad Q_{\mu\nu}(\bm{\mathcal{C}})=g_\mu^i(\bm{\mathcal{C}})g_\nu^i(\bm{\mathcal{C}})
\end{align}
can be written as
\begin{align}
\label{lang1}
\dot{\mathcal{C}}_\mu=-(Q_{\mu\nu}+L^a_{\mu\nu})\partial_{\bm{\mathcal{C}}_{{\nu}}}\phi+\partial_{\bm{\mathcal{C}}_\nu}L^a_{\mu\nu}+g_\mu^i(\bm{\mathcal{{{C}}}})\xi_i
\end{align}
%For brevity we define $d_\alpha(\bm{\mathcal{C}})\equiv Q_{\alpha\beta}(\bm{\mathcal{C}})\partial_{\bm{\mathcal{C}}_\beta}\phi$ and $r_\alpha(\bm{\mathcal{C}})\equiv J_\alpha(\bm{\mathcal{C}})-\alpha(\bm{\mathcal{C}})= L^a_{\alpha\beta}\partial_{\bm{\mathcal{C}}_\beta}\phi+\partial_{\bm{\mathcal{C}}_\beta}L^a_{\alpha\beta}$, which are like dissipative and reactive part respectively of current in equilibrium system. 
with antisymmetric {couplings} $L^a_{\mu\nu}\equiv F_{\mu\nu}e^{-\phi}$, where $F_{\mu\nu}$ itself is antisymmetric and defined by
\begin{align}
\label{asm}
r_{{\mu}} P_0(\bm{\mathcal{C}})\equiv\partial_{\mathcal{C}_\nu}F_{\mu\nu} \,.
\end{align}
{We now clarify this notation.}
{First}, $\phi\equiv \ln P_0(\bm{\mathcal{C}})$ is the logarithm of the
steady-state solution $P_0(\bm{\mathcal{C}})$ of the associated 
Fokker--Planck equation corresponding to Eq.~(\ref{lang}), namely
\begin{align}
\label{fpe}
\partial_tP(\bm{\mathcal{C}},t)=-\partial_{\mathcal{C}_\mu}\!\left[J_\mu(\bm{\mathcal{C}})
P(\bm{\mathcal{C}},t)-Q_{\mu\nu}(\bm{\mathcal{C}})\partial_{\mathcal{C}_\nu}P(\bm{\mathcal{C}},t)\right]
\end{align}
Following the equilibrium paradigm~\cite{chaikin1995principles}, the deterministic flux $J_\mu(\bm{\mathcal{C}})$ is broken into ``dissipative'' and ``reactive'' contributions, $d_\mu(\bm{\mathcal{C}})$ and $r_\mu(\bm{\mathcal{C}})\equiv J_\mu(\bm{\mathcal{C}})-{{d}}_\mu(\bm{\mathcal{C}})$, respectively, where the former has the explicit form
\begin{align}
\label{dis}
d_\mu(\bm{\mathcal{C}})=Q_{\mu\nu}(\bm{\mathcal{C}})\partial_{\bm{\mathcal{C}}_\nu}\phi
\end{align}
%(\ref{dis}), (\ref{fpe}) and definition of steady state distribution implies that in the steady state
%
%\begin{align}
%\label{div}
%\partial_{\mathcal{C}_\beta}[r_\beta P_0(\bm{\mathcal{C}})]=0
%\end{align}
Using Eq.~(\ref{asm}) and the definition $L^a_{\mu\nu}\equiv F_{\mu\nu}e^{-\phi}$, the reactive current can be cast into the explicit form
\begin{align}
r_\mu(\bm{\mathcal{C}})=L^a_{\mu\nu}\partial_{\bm{\mathcal{C}}_\nu}\phi+\partial_{\bm{\mathcal{C}}_\nu}L^a_{\mu\nu}
\end{align}

The point we want to make here is that the structure of the dissipative and reactive terms in the equilibrium and equilibrium-like description is exactly the same, i.e. Eqs.~(\ref{lang1}) and (\ref{noi1}) are structurally identical to Eqs.~(\ref{eq:genlang}) and  (\ref{eq:gennoise}). The identification of the symmetric coupling with the noise strength is also present in both {cases}, establishing an FDR of  the first {kind}, as noted in Ref.~\cite{eyink1996hydrodynamics}. 
One very important difference is that, unlike in equilibrium, the antisymmetric coupling is not obliged to originate from a Poisson bracket, in the generic case. It can however easily be seen that the generic antisymmetric coupling, $\mathbf{L^a}$ boils down to a Poisson bracket $\bm{\mathcal{W}}$, in the equilibrium limit, where $\phi=-H$. 
Then, the antisymmetric coupling is $\bm{\mathcal{W}}e^{\phi}e^{-\phi}$, and equating this with $\mathbf{L^a}$ gives
\begin{align}
\label{equ}
F_{\mu\nu}=\mathcal{W}_{\mu\nu}e^{-H}
\end{align}
Taking {its} derivative and using (\ref{asm}), we then find
\begin{align}
r_\nu=\mathcal{W}_{\mu\nu}\partial_{\mathcal{C}_\mu}H-\partial_{\mathcal{C}_\mu}\mathcal{W}_{\mu\nu} \,,
\end{align}
consistent with the equilibrium formalism~\cite{chaikin1995principles,ma1975sk}.

In summary, Eqs.~(\ref{eq:genlang}) and (\ref{lang1}) have an identical form, and the symmetry of the coupling coefficient with respect to an interchange of its indices is also the same. However, when the dynamics is governed by Eq.(\ref{eq:genlang}), the system is in equilibrium, whereas Eq.~(\ref{lang1}) can describe both equilibrium as well as nonequilibrium dynamics. This prompts the question where is the nonequilibrium condition hidden in Eq.~(\ref{lang1})? It is clear from the above discussion that an explicit description of a nonequilibrium dynamics by Eq.~(\ref{lang1}) requires the knowledge of its steady-state distribution. In general, the latter will be very difficult to find. Therefore, we study in the following an exactly solvable linear system to provide explicit answers to the theoretical questions raised and to illustrate the general statements promised in the Introduction \ref{intro}.

\section{A solvable model}
\label{sec3}
As described above, for a given (effective) {{Hamiltonian}} and noise correlation matrix  in an equilibrium system, the reactive and dissipative terms come out naturally with the correct time reversal signatures. But things get more complicated once the system is out of equilibrium. Here we discuss the linear case, which can be solved exactly, to see how the classification of the terms as reactive and dissipative looses meaning. Our starting point is the following set of coupled linear equations.
\begin{equation}
\label{modfmom}
\dot{\alpha} =-\frac{1}{\Gamma}\partial_{\alpha} H +{\tilde{\upsilon}}\partial_\beta{H}+\xi_\alpha(t)
\end{equation}
\begin{equation}
\label{modfaux}
\dot{\beta} =-{\tilde{\upsilon}}\partial_{\alpha} H-\frac{1}{\gamma} {\partial_\beta {H}}+{\xi}_\beta(t)
\end{equation}
associated with the quadratic Hamiltonian 
\begin{align}
\label{hami}
H=\frac{1}{2}K \alpha^2+\frac{1}{2}k\beta^2
\end{align}
where $K$ and $k$ are positive stiffness constants; {$\tilde{\upsilon}$ is a positive constant}.
The Markovian noise correlation matrix shall be given by
\begin{equation}
\label{Noi}
\mathbf{D}=2\begin{pmatrix}\frac{1}{\Gamma}&&0\\ 0 &&\frac{1}{\gamma}\end{pmatrix} \delta(t-t')
\end{equation}
with positive mobilities $1/\Gamma$ and $1/\gamma$.
The antisymmetric couplings can be assumed to be derived from a Poisson bracket 
 $\{\alpha,\beta\} =\tilde{\upsilon}$ \cite{dzyaloshinskii1980poisson}.  Since all  couplings are constant and hence even under time reversal, the equations describe equilibrium dynamics only when $\alpha$  and $\beta$ have opposite signature under time reversal \cite{casimir1945onsager,de2013non}.  Then the equilibrium distribution of the variables following this set of equations is proportional to $e^{-H}$. The discussion can easily be extended to the case where coupling constants are odd under time reversal (like magnetic fields), but this does not add anything new to the physics.
% Same steady state distribution can also be achieved when both $\alpha(\mathbf{x})$ and $\beta(\mathbf{x})$ are even or odd under time reversal and the coupling constant $\upsilon'$ is odd under time reversal (can be assumed to depend on magnetic field). 
Note that the choice of the Poisson brackets and noise covariance does still not uniquely fix the form of the Hamiltonian and the associated equations of motion~\cite{dzyaloshinskii1980poisson,stark2005poisson}.
But the remaining freedom does not allow to breach the equilibrium structure.

There are multiple ways to take Eqs.(\ref{modfmom}) and (\ref{modfaux})  out of equilibrium. Here we choose  to make the dynamics of $\beta$ autonomous, i.e., independent of  
$\alpha$, for which the dependence on $\beta$ is retained. 
The reciprocity of  the mutual forces is thereby broken  \cite{casimir1945onsager,de2013non}, the Hamiltonian structure is lost, and the nonequilibrium equations read
\begin{equation}
\label{modnmom}
\dot{\alpha} =-\frac{1}{\Gamma}\partial_{\alpha} H +{\tilde{\upsilon}}\partial_\beta{H}+\xi_\alpha
\end{equation}
\begin{equation}
\label{modnaux}
\dot{\beta} =-\frac{1}{\gamma} {\partial_\beta {H}}+{\xi}_\beta
\end{equation}
{Notice that these equations, in contrast to Eqs.~(\ref{modfmom}), (\ref{modfaux}), always correspond to nonequilibrium dynamics irrespective of any time reversal signature of $\alpha$ and $\beta$.}
%\SR{Note that although (\ref{modfmom}) and (\ref{modfaux}) are in equilibrium only if $\alpha$ and $\beta$ have opposite parity under $\mathcal{T}$ but (\ref{modnmom}) and (\ref{modnaux}) are nonequiilbrium for any parity of dynamical variables under $\mathcal{T}$.}
We keep the original noise covariance matrix $\mathbf{D}$ as in Eq.~\eqref{Noi}.
%\begin{equation}
%\label{NOI}
%{\mathbf{D}}=\begin{pmatrix}\frac{1}{\Gamma}&&0\\ 0 &&\frac{1}{\gamma}\end{pmatrix}\delta(t-t')
%\end{equation}
Equations (\ref{modnmom}) and (\ref{modnaux}) can then still be written in the equilibrium-like form of Eq.~(\ref{lang1}), 
albeit with an effective nonequilibrium Hamiltonian $\Phi$, as shown in App. \ref{appa}, and a new (apparently) ``reactive'' coupling.
\begin{equation}
\label{modf1mom}
\dot{\alpha} =-\frac{1}{\Gamma}\partial_{\alpha} \Phi +\frac{{\upsilon\Gamma}}{K\gamma+k\Gamma}\partial_\beta{\Phi}+\xi_\alpha
\end{equation}
\begin{equation}
\label{modf1aux}
\dot{\beta} =-\frac{{\upsilon\Gamma}}{K\gamma+k\Gamma}\partial_{\alpha} \Phi-\frac{1}{\gamma} {\partial_\beta {\Phi}}+{\xi}_\beta
\end{equation}
where $\upsilon=k\tilde{\upsilon}$ and
\begin{widetext}
\begin{equation}
\label{pot1}
\Phi=\frac{(K\gamma+k\Gamma)((K^2\alpha^2+({\upsilon\Gamma})^2\beta^2+K\beta(-2{\upsilon\Gamma} \alpha+k\beta))\gamma+k(K\alpha^2+k\beta^2)\Gamma)}{2(K^2\gamma^2+(({\upsilon\Gamma})^2+2kK)\gamma\Gamma+k^2\Gamma^2)}
\end{equation}
\begin{equation}
\label{for1}
\partial_{\alpha} \Phi=\frac{(K\gamma+k\Gamma)((K^2\alpha-K{\upsilon\Gamma} \beta)\gamma+kK\Gamma \alpha)}{(K^2\gamma^2+(({\upsilon\Gamma})^2+2kK)\gamma\Gamma+k^2\Gamma^2)}
\end{equation}
\begin{equation}
\label{for2}
\partial_\beta{\Phi}=\frac{(K\gamma+k\Gamma)((({\upsilon\Gamma})^2\beta+K(-{\upsilon\Gamma} \alpha+k\beta))\gamma+k^2\beta\Gamma)}{(K^2\gamma^2+(({\upsilon\Gamma})^2+2kK)\gamma\Gamma+k^2\Gamma^2)}
\end{equation}
\end{widetext}
{Recall from} Sec.~\ref{sec1} that equilibrium requires ``reactive'' and ``dissipative'' terms to have the same and opposite time-reversal symmetry as $\dot{\alpha}$ and $\dot{\beta}$, respectively. {If} $\alpha$ and $\beta$ have opposite signature under time reversal, the effective NH $\Phi$ in Eq.~(\ref{pot1}) is not invariant under time reversal, because it contains terms like $\alpha\beta$. {And if the time reversal signature of $\alpha$ and $\beta$ is the same (say even), the NH is even under time reversal. Which implies that the NH need not be time reversal invariant for nonequilibrium dynamics, also see Refs.~\cite{ferretti2022signatures,dal2021fluctuation}. An interesting observation, easily gleaned from Eqs.~(\ref{for1}) and (\ref{for2}) is that, in either case the  ``reactive'' and ``dissipative'' terms no longer exhibit the time reversal signature required for thermal equilibrium, which is a manifestation of nonequilibrium.} { We discuss this in more detail in the next paragraph.} {Although we have shown this feature explicitly only for the liner system, it is important to note that it needs to hold generally, also for nonlinear nonequilibrium systems}, since otherwise thermal equilibrium pertains. As a result, unlike the situation in thermal equilibrium, where only the dissipative term is responsible for entropy production during the evolution towards steady state, now both the ``dissipative'' and ``reactive'' terms contribute. Furthermore, while the ``dissipative'' fluxes and the divergence of the ``reactive'' fluxes vanish {in} the steady state \cite{graham1977covariant,eyink1996hydrodynamics}, as {in} equilibrium, the ``reactive'' flux now keeps producing ``housekeeping'' heat and entropy even after the steady state has been attained.  {The presence of housekeeping entropy production} is, of course, a necessary signature of any nonequilibrium steady state (NESS), which underscores the necessity of the condition for a system out of equilibrium.

\emph{Nonequilibrium Hamiltonian and detailed balance:} From  the explicit Eqs.~(\ref{modf1mom}), (\ref{modf1aux}), (\ref{for1}), (\ref{for2}) it is clear that one cannot restore the detailed balance required for equilibrium by assigning any other time reversal signature to the dynamical variables $\alpha$ and $\beta$. For detailed balance to hold, say, if both $\alpha$ and $\beta$ are even under time reversal, the ``reactive'' term should be odd, which is simply not the case for Eqs.~(\ref{for1}) and (\ref{for2}) . Similarly, as one easily convinces oneself, any other combination of time reversal signatures  for the variables will also not give rise to the required equilibrium symmetries for the  ``reactive'' and ``dissipative'' terms. Therefore, it is not possible to restore detailed balance even though one knows the NH $\Phi$ exactly~\cite{graham1971generalized}. That this is so, inevitably follows from the underlying breaking of reciprocity in {Eqs.~(\ref{modnmom}), (\ref{modnaux})}, which is somewhat masked by the splitting of the fluxes into nominally (but not literally) ``dissipative'' and ``conservative'' parts \cite{casimir1945onsager} in Eqs.~ (\ref{modf1mom}),(\ref{modf1aux}). 

{Generally speaking, it is always possible to construct a set of equilibrium Langevin equations for a $\mathcal{T}$-even steady state distribution but impossible otherwise \cite{risken1996fokker}. {Here is a formal way to see this.} The general condition for detailed balance (when both $\mathcal{T}$ even and odd variable are present) is \cite{risken1996fokker}
\begin{equation}
P_0(\bm{\mathcal{C}})w(\bm{\mathcal{C}}\rightarrow \bm{\mathcal{C'}})=P_0(\mathcal{T}\bm{\mathcal{C'}})w(\mathcal{T}\bm{\mathcal{C'}}\rightarrow \mathcal{T}\bm{\mathcal{C}})
\end{equation}
It can be telescoped for a chain of configurations:
\begin{equation}
\begin{aligned}
P_0(\bm{\mathcal{C}}_0) w(\bm{\mathcal{C}}_0\rightarrow \bm{\mathcal{C}}_1)\dots w(\bm{\mathcal{C}}_{n-1}\rightarrow \bm{\mathcal{C}}_n)=\\P_0(\mathcal{T}\bm{\mathcal{C}}_n)w(\mathcal{T}\bm{\mathcal{C}}_n\rightarrow \mathcal{T}\bm{\mathcal{C}}_{n-1})\dots w(\mathcal{T}\bm{\mathcal{C}}_1\rightarrow \mathcal{T}\bm{\mathcal{C}}_0)
\end{aligned}
\end{equation}
In particular, for cyclic state chages, i.e. $\bm{\mathcal{C}}_n=\bm{\mathcal{C}}_0$, 
\begin{equation}
\label{ndb}
\frac{w(\bm{\mathcal{C}}_0\rightarrow \bm{\mathcal{C}}_1)\dots w(\bm{\mathcal{C}}_{n-1}\rightarrow \bm{\mathcal{C}}_0)}{w(\mathcal{T}\bm{\mathcal{C}}_0\rightarrow \mathcal{T}\bm{\mathcal{C}}_{n-1})\dots w(\mathcal{T}\bm{\mathcal{C}}_1\rightarrow \mathcal{T}\bm{\mathcal{C}}_0)}=e^{[\phi(\bm{\mathcal{\mathcal{T}C}}_0)-\phi(\bm{\mathcal{C}}_0)]}
\end{equation}
For a Markov process defined with transition rates $w(\bm{\mathcal{C}}\rightarrow \bm{\mathcal{C'}})$, the entropy production in the heat bath is then given by \cite{evans1993probability,evans2002fluctuation,
derrida2007non}
\begin{equation}
\label{ment}
S=k\,\ln\frac{w(\bm{\mathcal{C}}\rightarrow \bm{\mathcal{C'}})}{w(\mathcal{T}\bm{\mathcal{C'}}\rightarrow \mathcal{T}\bm{\mathcal{C}})}
\end{equation}
Combining Eqs.~(\ref{ndb}) and (\ref{ment}) confirms that any $\mathcal{T}$-odd terms in the NH $\phi$ and the steady state distribution $\propto e^{\phi}$ give rise to the entropy production $S=k[\phi(\bm{\mathcal{\mathcal{T}C}}_0)-\phi(\bm{\mathcal{C}}_0)]$. {Of course, there can be further contributions on the top of this}. Another formal method illustrating this point is discussed in Refs.~\cite{ferretti2022signatures,dal2021fluctuation}.}

\emph{Symmetric versus antisymmetric coupling:}
To achieve an equilibrium-like form, the symmetric couplings {in the dynamical equations} are  required to  equal those in the noise covariance matrix. If two sets of dynamical equations have the same dissipative couplings and thus also the same noise covariances, they can  be understood as coupled to the same heat bath. In an equilibrium system with an effective Hamiltonian structure, the antisymmetric coupling is also predetermined, namely by the Poisson brackets between the dynamical variables.  In the case of coarse grained (field) variables, the antisymmetric couplings can either be calculated from the Poisson brackets of a microscopic model or written down phenomenologically, based on the knowledge of symmetries \cite{stark2005poisson,dzyaloshinskii1980poisson}.  However, far from equilibrium, an origin of the ``reactive'' couplings from Poisson brackets is no longer guaranteed. Then, similarly as for the symmetric couplings, one may wonder about the physical implications (if any) if two sets of equations share the same ``reactive'' couplings.

As an example, consider Eq.~(\ref{modnmom}) without the noise term for  $\alpha$. Then the corresponding equilibrium-like formulation of Eqs.~(\ref{modnmom}), (\ref{modnaux}) takes the new form 
\begin{equation}
\label{modgmom}
\dot{\alpha} =\frac{{\upsilon\Gamma}}{K\gamma+k\Gamma}\partial_\beta{\phi}
\end{equation}
\begin{equation}
\label{modgaux}
\dot{\beta} =-\frac{{\upsilon\Gamma}}{K\gamma+k\Gamma}\partial_{\alpha} \phi-\frac{1}{\gamma} {\partial_\beta {\phi}}+{\xi}_\beta
\end{equation}
which differs from Eqs.~(\ref{modf1mom}) and (\ref{modf1aux}), and where also
\begin{equation}\label{phi_def}
\phi=\frac{K\gamma+k\Gamma}{2({\upsilon\Gamma})^2\gamma\Gamma}((K\alpha-{\upsilon\Gamma} \beta)^2\gamma+kK\Gamma \alpha^2)
\end{equation}
is a new NH, not equal to $\Phi$.
Since we suppressed the noise for $\alpha$, the noise correlation changes to
\begin{equation}
{\mathbf{D}}=2\begin{pmatrix}0&&0\\ 0 &&\frac{1}{\gamma}\end{pmatrix}\delta(t-t')
\end{equation}
Notice, that the antisymmetric coupling coefficient in Eqs.~(\ref{modgmom}) and (\ref{modgaux}) is however the same as in Eqs.~(\ref{modf1mom}) and (\ref{modf1aux}). While, even far from equilibrium, for two sets of equations to have identical symmetric couplings implies that they have the same (virtual) thermostat, possible implications of identical ``reactive'' coupling coefficients, as in the above example, are less clear and might deserve further study.

\section{Active Ornstein--Uhlenbeck Particles (AOUP)}
\label{sec4}
The active Ornstein--Uhlenbeck process is a nonequilibrium variant of the well-known equilibrium Ornstein--Uhlenbeck process (the stochastic harmonic oscillator) \cite{uhlenbeck1930theory}. It can be interpreted as the equation of motion of an active Brownian particle or microswimmer \cite{sandford2017pressure}. The particle coordinate is given by $X(t)$ and a nominal (autonomous swimming is actually a force-free motion) propulsion force  is given by $x(t)$. 
The latter defines the  direction of swimming, which is a stochastic variable, whereas the stochasticity of the particle's {center-of-mass} coordinate is omitted, since it is negligible  compared to the systematic swimming motion, at the relevant late times.
Here we consider the one-dimensional case, as in Ref.~\cite{sandford2017pressure}, which can easily be generalized to higher spatial dimensions \cite{fodor2016far}. The equation of motion is
\begin{equation}
\label{mod2mom}
\Gamma \dot{X} =-{\partial_X H} +{ x}
\end{equation}
\begin{equation}
\label{mod2aux}
\gamma \dot{x} =- x+{\xi}_x
\end{equation}
where $H$ represents an external potential. The case  with a harmonic confinement potential $H=1/2KX^2$ is studied in detail in Ref.~\cite{sandford2017pressure}. It corresponds exactly to Eqs.~(\ref{modnmom}) and (\ref{modnaux}) without a noise term for $\alpha$, if $\alpha$ and $\beta$ are identified with $X$ and $x$, respectively. The equilibrium-like structure is the one provided  in  Eqs.~(\ref{modgmom}), (\ref{modgaux}).  Recall that
 the FPE corresponding to a generic Langevin equation is solved {for the stationary state} by setting the divergence of the probability flux  to zero. When the steady state corresponds to a  thermal equilibrium, additionally the  dissipative part of the flux itself has to vanish identically, to avoid any spurious entropy production. 
%and can be useful to give insight into the nature of the problem. WHAT ?
Now, for a nonequilibrium Langevin equation written in equilibrium-like form, like the set of Eqs.~(\ref{modgmom}), (\ref{modgaux}) in case of Eqs.~(\ref{mod2mom}) and (\ref{mod2aux}), the steady-state solution formally looks like a Boltzmann equilibrium (while it is not). Thanks to the equilibrium-like formulation, the nominally dissipative part of the flux, which tracks the relaxation to the steady state, still vanishes in the steady state \cite{graham1977covariant,eyink1996hydrodynamics}, so that only the nominally reactive flux can account for the housekeeping heat and entropy production. This is how the splitting of the total flux into nominally ``reactive'' and ``dissipative'' parts is still useful and gives insight into the nature of entropy production in such situations far from equilibrium.  {For a diferent kind of splitting of the fluxes on the basis of their parity under $\mathcal{T}$ we refer the reader to Refs.~\cite{ferretti2022signatures,dal2021fluctuation} .}
For a discussion of the explicit result for the entropy production of the above AOUP system, see Ref.~\cite{sandford2017pressure,dadhichi2018origins,caprini2019entropy}. {In the following section, 
our emphasis is more on its general structure.}

\section{Entropy production}
\label{sec5}
{ The primary aim of this section is to show how the $\mathcal{T}$-odd and $\mathcal{T}-$even parts of the NH contribute to the entropy production. For concreteness, our discussion is based on the above model Eqs. (\ref{modf1mom}) and (\ref{modf1aux}), but the results are more general.}

{The steady-state entropy production rate is defined as \cite{lebowitz1999gallavotti}
\begin{equation}
\label{defent}
\sigma=\lim\limits_{\tau \to \infty}\frac{1}{\tau}S,\quad S=\langle \ln({P}/{P^R})\rangle
\end{equation}
which is recognised as the Kullback–Leibler (KL) divergence \cite{kullback1951information}. It measures the distinguishability of the probability weight, $P$, associated with a path $\{\alpha(t),\beta(t)\}_{0\le t\le\tau}$ and the weight $P^R$ for the time-reversed path. The angular brackets denote an average over noise realisations. However, under suitable ergodicity assumptions, which we implicitly use throughout the paper, the average over noise realizations can be replaced by the time average over a single infinitely long noise realisation. Therefore, the angular brackets can be dropped under (or exchanged for) time avereages \cite{fodor2016far,nardini2017entropy}. As discussed in App.~\ref{appb} \cite{markovich2021thermodynamics,lau2007state,cugliandolo2017rules}, the trajectory probability  $P= e^{-A}$ can be expressed in terms of the action $A$,  which for (\ref{modf1mom}) and (\ref{modf1aux}) in Stratonovich convention reads}
{
\begin{equation}
\begin{aligned}
A=&\int dt \,\frac{1}{4}\Bigg[\Gamma\left(\dot{\alpha} +\frac{1}{\Gamma}\partial_{\alpha} \Phi -\frac{{\upsilon\Gamma}}{K\gamma+k\Gamma}\partial_\beta{\Phi}\right)^2\\&+\gamma\left(\dot{\beta} +\frac{{\upsilon\Gamma}}{K\gamma+k\Gamma}\partial_{\alpha} \Phi+\frac{1}{\gamma} {\partial_\beta {\Phi}}\right)^2\!\!\!-\frac{2}{\Gamma}\partial_\alpha^2\Phi-\frac{2}{\gamma}\partial_\beta^2\Phi\Bigg]
\end{aligned}
\end{equation}
The entropy production rate (\ref{defent}), expressed as a function of the action, is then
\begin{equation}
\sigma=\lim\limits_{\tau \to \infty}\frac{A^R-A}{\tau}
\end{equation}
where $A^R\equiv \mathcal{T}A$.
For the case when both noises have equal strength, i.e. $\Gamma=\gamma$, {and $\alpha$ and $\beta$ have opposite parity under $\mathcal{T}$},  App.~\ref{appf} gives
\begin{equation}
\label{entpro}
\begin{split}
\sigma= -\lim\limits_{\tau \to \infty}\frac{1}{\tau}\int dt \Bigg(\frac{d\Phi_s}{dt}+M\left(\partial_\alpha \Phi_a \partial_\alpha\Phi_s+\partial_\beta \Phi_a \partial_\beta\Phi_s\right)\\+\frac{{\upsilon}\gamma}{K+k}\left(\dot{\beta}\partial_{\alpha} \Phi_a-\dot{\alpha}\partial_\beta{\Phi}_a\right)-\frac{1}{\gamma}\left(\partial_\alpha^2\Phi_a+\partial_\beta^2\Phi_a\right)\Bigg)\\ =
-\lim\limits_{\tau \to \infty}\frac{1}{\tau}\Bigg[\Delta\Phi_s+\int dt \Bigg(M\partial_{\bm{\mathcal{C}}} \Phi_a\cdot \partial_{\bm{\mathcal{C}}} \Phi_s\hspace*{2cm}\\+\frac{{\upsilon}\gamma}{K+k}\partial_{\bm{\mathcal{C}}} \Phi_a\wedge\dot{\bm{\mathcal{C}}}-\frac{1}{\gamma}\partial_{\bm{\mathcal{C}}}^2\Phi_a\Bigg)\Bigg]
\end{split}
\end{equation}
{Here, $M$ is a positive constant given in App.~\ref{appf}}; $\Phi_s$ and $\Phi_a$ are the $\mathcal{T}$-even and $\mathcal{T}$-odd parts of the NH respectively. The first term, $\Delta\Phi_s=\Phi_s(\tau)-\Phi_s(0)$ is simply the difference between the final and initial values of the $\mathcal{T}$-even part of the NH. Since $\Delta\Phi_s$ is always finite, it contributes  transiently and vanishes in the steady state due to the division by ${\tau \to \infty}$. Note that for an equilibrium system, the Hamiltonian is always $\mathcal{T}$-even. Therefore the last three terms vanish in this case, leading to zero entropy production in the steady state. The second term changes sign depending on whether the forces originating from $\mathcal{T}$-even and $\mathcal{T}$-odd parts of the NH oppose or align with each other. The latter force is of  nonequilibrium  nature. The  second last term with the wedge product (here equivalent to the cross product in two dimension) is reminiscent of  classical Hamiltonian dynamics in phase space, which is perpendicular to the energy gradients due to the symplectic Hamiltonian structure. Yet, it is contributing to the entropy production, because $\Phi_a$ is $\mathcal{T}$-odd. The last term originates from the Jacobian in the Stratonovich discretization convention, see App.~\ref{appb}.} 

%\begin{figure}[t]
%\includegraphics[width=6cm]{pic9.pdf}
%\centering
%\caption{Schamatic diagram: Only the iso-potential component of the motion is responsible for the entropy production in the steady state when $\alpha$ and $\beta$ have even parity under time reversal. }
%\label{fig1}
%\end{figure}

{Inserting the explicit forms of $\Phi_s$ and $\Phi_a$ in the above expression gives
\begin{equation}\label{eq:harmonic}
\begin{aligned}
%\begin{split}
\sigma & =
-\lim\limits_{\tau \to \infty}\frac{1}{\tau}\Bigg[\Delta\Phi_s-\int dt \Bigg(M'\a\b-N'\frac{d(\a^2-\b^2)}{dt}\Bigg)\Bigg]\\ & = M'\langle\a\b\rangle
%\end{split}
\end{aligned}
\end{equation}
Here, $M'$ and $N'$ are positive constants given in App.~\ref{appf}. We have used the ergodic assumption and
the fact that $\Delta\Phi_s$ and $\Delta(\a^2-\b^2)$ are finite. Since our NH is a quadratic function of the dynamical variables, the dot product of the forces due to $\mathcal{T}$-even and $\mathcal{T}$-odd parts of the NH is the only entropy producing term in the steady state (for a more general NH, all terms would contribute). The corresponding correlator $\langle\a\b\rangle=\nu\gamma K/k(K\gamma+k\Gamma)$ is easily calculated \cite{dadhichi2018origins}.  Obviously, the steady state entropy production is always positive.}

{In the more general case where the noise strengths differ, $\Gamma^{-1} \neq \gamma^{-1}$, the  form (\ref{entpro}) is recovered, by a straightforward rescaling
$\alpha\rightarrow {\alpha}/{\sqrt{\Gamma}}$ and $\beta\rightarrow {\beta}/{\sqrt{\gamma}}$ with new prefactors (App.~\ref{appf})
\begin{equation}
\label{entpro1}
\begin{split}
A-A^R = \int dt \Bigg(\frac{d\Phi'_s}{dt}+ \tilde{M}\partial_{\bm{\mathcal{C}}}\Phi'_a\cdot \partial_{\bm{\mathcal{C}}}\Phi'_s \\ +\frac{{\upsilon\Gamma}\sqrt{\Gamma\gamma}}{K\gamma+k\Gamma}\partial_{\bm{\mathcal{C}}} \Phi'_a\wedge\dot{\bm{\mathcal{C}}}-\partial_{\bm{\mathcal{C}}}^2\Phi'_a\Bigg)
\end{split}
\end{equation}
where $\Phi'\equiv \Phi({\alpha}/{\sqrt{\Gamma}},{\beta}/{\sqrt{\gamma}})$, and $\tilde{M}$ is a constant. {The above results can be generalized to a generic noise covariance matrix, as long as it is positive definite, if the dynamical variables have a definite parity under $\mathcal{T}$  in the new coordinates \cite{ferretti2022signatures}. If all dynamical variables have the same this is automatically guaranteed.} It is important to note that the splitting of the entropy production rate found in Eqs.~(\ref{entpro}) and (\ref{entpro1}) is independent of the explicit form of  the NH and hence holds for generic nonlinear systems,  as long as noise is additive with no cross correlation and the antisymmetric coupling matrix is constant. Only the reduction to Eq.~(\ref{eq:harmonic}) is specific to the harmonic form of NH. {We discuss the entropy production rate for the more generic case in App.~\ref{appg}}. {A different interesting way of splitting entropy production is discussed in Refs.~\cite{ferretti2022signatures,dal2021fluctuation}.}
} 

{
As shown in Eq.~(\ref{genent}), when both  dynamical variables ($\alpha$ and $\beta$)  are $\mathcal{T}$-even, so that $\Phi=\Phi_s$, the entropy production rate is proportional to $\partial_{\bm{\mathcal{C}}} \Phi'\wedge\dot{\bm{\mathcal{C}}}$, {calculated explicitly in Ref.~\cite{dadhichi2018origins,b}}. It implies that, in the steady state, only the motion perpendicular to the gradient of the NH is responsible for the entropy production, {quite unlike equilibrium situation}. Although Eqs.~(\ref{modf1mom}) and (\ref{modf1aux}) have the structure of  equilibrium dynamics, the ``reactive" terms have the opposite time reversal signature compared to what is required in  equilibrium. {The dynamics given by (\ref{modf1mom}) and (\ref{modf1aux}), in the steady state, drives the system to the minimum of the NH, see App.~\ref{appl}, where the systematic forces vanish. But the noise keeps kicking the system out of the minimum, leading to said entropy production via what could be called ``active/dissipative" (external)  fluctuations. This situation is again  unlike equilibrium, since equilibrium thermal fluctuations are not externally driven nor do they contribute to dissipation.} {The origin of this discrepancy is the following. In equilibrium, the path leading to a fluctuation is the time reversal of its relaxation path \cite{onsager1953fluctuations}, whereas spontaneous nonequilibrium  fluctuations most likely follow a different trajectory, which is not the time reversed relaxation. For more on this interesting feature, which deserves further discussion, we refer to the  Refs.~\cite{bertini2001fluctuations,bertini2015macroscopic}.  }{It is also interesting to note that, in the case where the  variables have opposite signature under $\mathcal{T}$, the  entropy production rate has a contribution independent of the velocity $\dot{\bm{\mathcal{C}}}$, see Eq.~(\ref{entpro}). }
}

\section{Frenesy}
\label{sec6}
Graham \cite{graham1977covariant} extended the FDR of the second kind to the general (nonequilibrium) Fokker--Plank equation (\ref{fpe}) without invoking the notion of time-reversal symmetry. However, the quantity that provides the most natural  measure for the distance of a nonequilibrium system from thermal equilibrium, namely entropy production, is closely tied to the time reversal of a process and the dynamical variables involved. More recently, a new quantity called frenesy was introduced as an additional trait to characterize dynamics far from equilibrium 
\cite{maes2020frenesy,maes2020response}. {It accounts for some undirected (nonequilibrium) ``activity''.} Both quantities, {entropy production and frenesy,} are moreover closely related to the FDR \cite{harada2005equality,maes2020response}
 and hence natural concepts to be discussed in the context of the generalised FDR. 

As shown in App.~\ref{appb}, the action determines the relative weight of a stochastic trajectory. 
Frenesy is defined as its time-symmetric part, a measure of the escape rate from a state and a measure of 
undirected traffic~\cite{maes2020frenesy,maes2017non,maes2020response}. Frenesy also 
plays a key role in the non-linear response of an equilibrium state and, even {for} the linear response of non-equilibrium steady states \cite{maes2020response}. {As mentioned in the introduction, the NH acts as a Lyponov function for relaxations to steady states \cite{graham1977covariant} and also as entropy change upon adiabatic transitions between steady states \cite{hatano2001steady}}. For a given steady state, it is therefore natural to study frenesy in the context established above. {So let us again turn to the Eqs~(\ref{modf1mom}) and (\ref{modf1aux}), which can describe an AOUP with an additional noise acting onto the center of mass.} 

In situations far from equilibrium, for which there is in general no Hamiltonian, perturbing forces are usually added directly to the equation of motion. The extra frenesy of the AOUP  within this approach was already calculated before~\cite{maes2020response}. But in Graham's equilibrium-like formulation, a fundamentally different option arises. Due to presence of the NH, it is possible to add a work term depending {on} the perturbing force to the NH, which then yields entirely different equations of motion. If we follow this route for the AOUP model, and perturb its nonequilibrium Hamiltonian in Eq.~(\ref{pot1}) according to $\Phi\rightarrow \Phi-f_i(t)q_i$, where $q_1=\alpha$ and $q_2=\beta$, as in Ref.~\cite{graham1977covariant}, the resulting perturbed equation of motion is 
{
\begin{equation}
\label{fr1}
\dot{\alpha} =-\frac{1}{\Gamma}\partial_{\alpha} \Phi +\frac{{\upsilon\Gamma}}{K\gamma+k\Gamma}\partial_\beta{\Phi}+\frac{f_1}{\Gamma}-\frac{\upsilon\Gamma f_2}{K\gamma+k\Gamma}+\xi_\alpha
\end{equation}
\begin{equation}
\label{fr2}
\dot{\beta} =-\frac{{\upsilon\Gamma}}{K\gamma+k\Gamma}\partial_{\alpha} \Phi-\frac{1}{\gamma} {\partial_\beta {\Phi}}+\frac{{\upsilon\Gamma} f_1}{K\gamma+k\Gamma}+\frac{f_2}{\gamma}+{\xi}_\beta
\end{equation}}
{It is interesting to note that the perturbing force $f_1$, which couples only to $\alpha$ in the NH,  acts on both $\alpha$ and $\beta$ in the equation of motion, and \emph{vice-versa}  for $f_2$.
} And this is clearly the consequence of the antisymmetric coupling that comes with the Hamiltonian structure in this formulation. Usually, in an equilibrium system, 
we would understand this as a result of reactive dynamics originating from a Poisson bracket~\cite{chaikin1995principles}, 
but here we cannot rely on such a structure to exist. {Therefore even if the perurbing force conjugate to any single variable is omitted in the NH, the dynamics of this variable will still be directly perturbed.}
%\AM{Also note that, if the perturbing force coupling to $x$ is set to zero, i.e., $f_2=0$, 
%it is actually the dynamics of $x$ (and only $x$) that is directly perturbed. 
%The dynamics of $X$, which is couples directly to the perturbing force in the NH, 
%is then only affected indirectly, through $x$.
%As shown in App.~\ref{appc}, the dynamics governed by Eqs.~(\ref{pmod2mom}), (\ref{pmod2aux}) can be condensed into a single second-order differential equation for $X$ alone, 
%\begin{equation}
%\label{effe}
%\begin{aligned}
%\ddot{X} & =-\dot{X}\left(\frac{K}{\Gamma}+\frac{1}{\gamma}\right)-\frac{KX}{\gamma\Gamma}  -\frac{{f}_2}{\gamma(K\gamma+\Gamma)} \\ & - \frac{\dot{f}_2}{K\gamma+\Gamma}  + \frac{{f}_1}{\Gamma(K\gamma+\Gamma)}+\frac{f_2}{\Gamma\gamma}+\frac{\xi_x}{\Gamma}
%\end{aligned}
%\end{equation}}
%\AM{It is important to note that, in the absence of the perturbing forces, this is nothing but a damped harmonic oscillator driven by equilibrium noise. The step from Eqs.~(\ref{pmod2mom}), (\ref{pmod2aux}) to Eq.~(\ref{effe}) involves the elimination of the variable $x$, hence a ``coarse-graining'' over $x$ \cite{fodor2016far,dadhichi2018origins}. In App.~\ref{appc}, we show how to make the effective dynamics of $X$ equilibrium-like. Indeed, it results in zero entropy production, as long as the perturbing forces are absent, as was explicitly shown in Ref.~\cite{dadhichi2018origins}.}
Also, according to App.~\ref{appc}, the probability $P[\omega]$ of the trajectory $\omega=(X_s, s\in [0,t])$ is generally expressed in terms of the perturbed and unperturbed action $A$ and $A_0$, respectively, as
\begin{equation}
P[\omega]=e^{-A}=e^{-A_f(\omega)}P_\text{ref}[\omega]
\end{equation}
Here $P_\text{ref}[\omega]\equiv e^{-A_0(\omega)}$ is the probability in the absence of any perturbing forces. 
To leading order in the latter,  
{
\begin{equation}
\begin{aligned}
A_f=&\int ds\,\Bigg[\left(\frac{1}{\Gamma}+\frac{({\upsilon\Gamma})^2\gamma }{(K\gamma+k\Gamma)^2}\right)\left(\frac{f_1}{\Gamma}\partial_{\alpha}\Phi+\frac{f_2}{\gamma}\partial_{\beta}\Phi\right)\\&+\Gamma\dot{\alpha}\left(\frac{f_1}{\Gamma}-\frac{{\upsilon\Gamma} f_2}{K\gamma+k\Gamma}\right)+\gamma\dot{\beta}\left(\frac{{\upsilon\Gamma} f_1}{K\gamma+k\Gamma}+\frac{f_2}{\gamma}\right)\Bigg]
\end{aligned}
\end{equation}}
%\AM{\begin{equation}
%\begin{aligned}
%A_f&=\int_0^t\gamma\Gamma \left[\frac{1}{\gamma}\left(\frac{f_1}{\Gamma}-\frac{\upsilon\Gamma}{K\gamma+\Gamma}f_2\right)\left(\dot{X} +\frac{KX}{\Gamma} -\frac{\upsilon\Gamma x}{\Gamma}\right)\right]  \\&  +\left[\frac{1}{\Gamma}\left(\frac{{\upsilon\Gamma}}{K\gamma+\Gamma} f_1+\frac{f_2}{\gamma}\right)\left(\dot{x}+ \frac{kx}{\gamma} \right)\right]ds
%\end{aligned}
%\end{equation}
%}
The behavior of the action $A$ under a time reversal operation $\mathcal{T}$ is among the central characteristics of nonequilibrium systems\cite{markovich2021thermodynamics,dadhichi2018origins,maes2020frenesy}. And its antisymmetric and symmetric parts are closely related to the concepts of entropy and frenesy, respectively~\cite{maes2020frenesy}. Since we are interested in perturbations and fluctuations around a steady state, what matters is the $A_f$ part of the action, very much as discussed in Ref.~\cite{maes2020frenesy}; see also App.~\ref{appc}. {If $\alpha$ and $\beta$ behave like coordinates, then they are even under time reversal. The time-symmetric part of  $A_f$ is then} 
{
\begin{equation}
\begin{aligned}
A_f+\mathcal{T}A_f=&2\int ds\,\Bigg[\left(\frac{1}{\Gamma}+\frac{({\upsilon\Gamma})^2\gamma }{(K\gamma+k\Gamma)^2}\right)\mathbf{f}\cdot\partial_{\bm{\mathcal{C}}}\Phi'\Bigg]
\end{aligned}
\end{equation}
where, as above, a rescaling of  $\alpha\rightarrow {\alpha}/{\sqrt{\Gamma}}$ and $\beta\rightarrow {\beta}/{\sqrt{\gamma}}$; $\Phi'\equiv \Phi({\alpha}/{\sqrt{\Gamma}},{\beta}/{\sqrt{\gamma}})$ was employed and $\mathbf{f}$ is the perturbing force. For our particular form of $\Phi$:
\begin{equation}
\begin{aligned}
&A_f+\mathcal{T}A_f=\gamma\Gamma\int \Bigg[\frac{1}{\gamma}\left(\frac{f_1}{\Gamma}-\frac{{\upsilon\Gamma}}{K\gamma+\Gamma}f_2\right)\left(\frac{K\alpha}{\Gamma} -\frac{{\upsilon\Gamma} \beta}{\Gamma}\right)  \\&  +\frac{1}{\Gamma}\left(\frac{\nu}{K\gamma+\Gamma} f_1+\frac{f_2}{\gamma}\right)\left(\frac{k\beta}{\gamma} \right)\Bigg]ds
\end{aligned}
\end{equation}
}
%\AM{\begin{equation}
%\begin{aligned}
%A_f&-\mathcal{T}A_f =\int_0^t\gamma\Gamma \left[\frac{1}{\gamma}\left(\frac{f_1}{\Gamma}-\frac{\upsilon\Gamma}{K\gamma+\Gamma}f_2\right)\left(\dot{X} -\frac{\upsilon\Gamma x}{\Gamma}\right)\right]  \\&  +\left[\frac{1}{\Gamma}\left(\frac{\upsilon\Gamma}{K\gamma+\Gamma} f_1+\frac{f_2}{\gamma}\right)\left(\frac{kx}{\gamma} \right)\right]ds
%\end{aligned}
%\end{equation}
%}
The time-antisymmetric part
{
\begin{equation}
\begin{aligned}
&A_f-\mathcal{T}A_f=\\&2\int ds\,\Bigg[\Gamma\dot{\alpha}\left(\frac{f_1}{\Gamma}-\frac{{\upsilon\Gamma} f_2}{K\gamma+k\Gamma}\right)+\gamma\dot{\beta}\left(\frac{{\upsilon\Gamma} f_1}{K\gamma+k\Gamma}+\frac{f_2}{\gamma}\right)\Bigg]
\end{aligned}
\end{equation}
is not explicitly dependent on the NH.}

{If $\alpha$ and $\beta$ behave like coordinate and velocity, then they are even and odd under time reversal respectively. The time-symmetric part of  $A_f$ is then
\begin{equation}
\begin{aligned}
&A_f+\mathcal{T}A_f=\\&2\int ds\,\Bigg[\left(\frac{1}{\Gamma}+\frac{({\upsilon\Gamma})^2\gamma }{(K\gamma+k\Gamma)^2}\right)\left(\frac{f_1}{\Gamma}\partial_{\alpha}\Phi_s+\frac{f_2}{\gamma}\partial_{\beta}\Phi_a\right)\\&+\gamma\dot{\beta}\left(\frac{{\upsilon\Gamma} f_1}{K\gamma+k\Gamma}+\frac{f_2}{\gamma}\right)\Bigg]
\end{aligned}
\end{equation} 
} 
For our particular form of $\Phi$:
{\begin{equation}
\begin{aligned}
&A_f+\mathcal{T}A_f =2\int ds\, \Bigg[{K\alpha}\left(\frac{f_1}{\Gamma}-\frac{{\upsilon\Gamma}}{K\gamma+\Gamma}f_2\right) \\&  +\frac{1}{\Gamma}\left(\frac{{\upsilon\Gamma}}{K\gamma+\Gamma} f_1+\frac{f_2}{\gamma}\right)\dot{\beta}\Bigg]
\end{aligned}
\end{equation} 
}
Now, the antisymmetric action is
\begin{equation}
\begin{aligned}
&A_f-\mathcal{T}A_f=\\&2\int ds\,\Bigg[\left(\frac{1}{\Gamma}+\frac{({\upsilon\Gamma})^2\gamma }{(K\gamma+k\Gamma)^2}\right)\left(\frac{f_1}{\Gamma}\partial_{\alpha}\Phi_a+\frac{f_2}{\gamma}\partial_{\beta}\Phi_s\right)\\&+\Gamma\dot{\alpha}\left(\frac{f_1}{\Gamma}-\frac{{\upsilon\Gamma} f_2}{K\gamma+k\Gamma}\right)\Bigg]
\end{aligned}
\end{equation}
and, for our particular form of $\Phi$,
{
\begin{equation}
\begin{aligned}
A_f-\mathcal{T}A_f&=2\int\Gamma \Bigg[\left(\frac{f_1}{\Gamma}-\frac{\upsilon\Gamma}{K\gamma+\Gamma}f_2\right)\left(\dot{\alpha} -\frac{\upsilon\Gamma \beta}{\Gamma}\right)  \\&  +\frac{1}{\Gamma}\left(\frac{{\upsilon\Gamma}}{K\gamma+\Gamma} f_1+\frac{f_2}{\gamma}\right) \frac{k\beta}{\gamma}\Bigg]ds
\end{aligned}
\end{equation}
}
To compare these results with {the conventional expressions for entropy production and frenesy}, note that
the time derivatives can be interchanged between the coordinates and the perturbing forces using integration by parts.
Interestingly, the perturbing forces in Eqs.~(\ref{fr1}), (\ref{fr2}) appear in a different, inequivalent way
compared to the above mentioned case, when the perturbing forces are directly added to the equation of motion; see Refs.~\cite{maes2020response,dadhichi2018origins} and App.~\ref{appd}. As a result, the entropy and frenesy  contributions to the action {(and thus the explicit type of activity associated with them)} differ objectively from those discussed in Ref.~\cite{maes2020response}.

Finally for completeness, let us give an outlook onto the role of Graham’s FDR for the Harada—Sasa relation \cite{harada2005equality}.
The later states that the housekeeping heat
associated with a nonequilibrium steady-state manifests itself 
as a violation of the equilibrium FDR, obtained when a weak
perturbing force is introduced to the equation of motion. 
In the nonequilibrium steady-state, the difference between 
the correlation and response function grows proportionally 
with the dissipation rate.  As we have discussed, Graham’s
equilibrium-like reformulation of equations admitting a
nonequilibrium steady-state always admits a formal FDR
for perturbing forces added into the NH. Combining this
with the Harada—Sasa relation, one concludes that the
housekeeping dissipation also amounts to a measure of 
the discrepancy between the conventional linear response
and its formal counterpart obtained from the NH. 
In other words, the three metrics for quantifying the distance
to equilibrium (steady-state dissipation, FDR violation,
and the discrepancy between conventional and NH notions of 
linear response) are mutually consistent. In the same vein, it should
also be informative to keep track of the discrepancy between the corresponding
alternative notions of frenesy discussed above. 

\section{Conclusion}
\label{sec7}
In this paper we have considered nonequilibrium {Markovian Langevin equations}, where the potential  {corresponding to the logarithm of the steady-state distribution}  can formally take the role of a free energy or coarse-grained Hamiltonian, which we called the nonequilibrium Hamiltonian. {Exactly solving an explicit model }we find that { the NH does not need to be} a time-reversal invariant quantity, out of equilibrium. {Even if the existence of a steady state is exploited to rephrase nonequilibrium Langevin equations, so that they appear formally equivalent to those of an equilibrium system, ``reactive'' and ``dissipative'' terms  lack important symmetries required for detailed balance.} In particular, there are no Poisson-brackets underlying the ``reactive'' couplings.  In equilibrium Langevin equations, the reactive terms are reversible and confined to level surfaces of the Hamiltonian, so that they produce no heat and entropy. In the equilibrium-like {{Langevin}} equation, ``reactive'' terms are responsible for ``housekeeping'' entropy production that {maintains} the steady state and {prevents} its relaxation to equilibrium. {Far from equilibrium, a splitting of the NH into time-reversal even and odd parts is observed to be useful, since both parts contribute in different, physically transparent  ways to the entropy production.} { In some cases, the entropy production comes entirely from {what could be called} ``active/dissipative" fluctuations.}
We also discussed two non-equivalent but physically meaningful {instances of frenesy}, depending on whether the relevant perturbing forces are added directly to the equations of motion or rather introduced at the level of the nonequilibrium Hamiltonian. {As an outlook, we pointed out that the response corresponding to the later type of perturbation can lead to a variant of the Harada-Sasa relation.}
Finally,  it might be interesting to further investigate the physical significance of two sets of nonequilibrium Langevin equations having the same antisymmetric coupling in their equilibrium-like formulation. 

\section*{ACKNOWLEDGMENTS}
LPD was supported by the Alexander von Humboldt Foundation's Humboldt Research Fellowship for Postdoctoral Researchers. { We thank the referees for their valuable comments and for bringing the Refs.~\cite{ferretti2022signatures,dal2021fluctuation} to our attention.}

\begin{appendix}	
\section{Exact solution of the equation of motion}
\label{appa}
We show here how we derived (\ref{pot1}) in section \ref{sec3}. Writing (\ref{modnmom}) and (\ref{modnaux}) using (\ref{hami}) in the vectorial form
\begin{align}
\label{veceq}
\dot{\mathbf{q}}=\mathbf{M}\cdot\mathbf{q}+\bm{\xi}
\end{align}
where $\mathbf{q}=(\alpha, \beta)$,
$\mathbf{M}=\begin{pmatrix}-\frac{K}{\Gamma}&&{\upsilon}\\ 0 &&-\frac{k}{\gamma}\end{pmatrix}$
and $\langle\bm{\xi(t)}\bm{\xi(t')}\rangle=\begin{pmatrix}\frac{1}{\Gamma}&&0\\ 0 &&\frac{1}{\gamma}\end{pmatrix}\delta(t-t')$, the solution of \ref{veceq} can be written as
\begin{align}
\mathbf{q}=\int_{-\infty}^t\!\!\!\exp[\mathbf{M}(t-s)]\bm{\xi(s)}ds
\end{align}
The corresponding covariance matrix, $C(t,t')\equiv\langle\mathbf{q}(t)\mathbf{q}(t')\rangle$, is given by
{{
\begin{equation}
\begin{aligned}
&\mathcal{C}(t,t')=\mathcal{C}(|t-t’|)=\\& \int_{-\infty}^t\int_{-\infty}^{t'}\!\!\!\!ds\,ds'\exp[\mathbf{M}(t-s)]\langle{\bm\xi(s)}{\bm\xi(s')}\rangle\exp[\mathbf{M}^T(t'-s')]
\end{aligned}
\end{equation}
}}
Since (\ref{veceq}) is a linear equation driven by a Gaussian noise, its steady state distribution is $\rho\propto\exp[-\Phi(\alpha,\beta)]$, where
\begin{widetext}
\begin{equation}
\Phi(\alpha,\beta)=\mathbf{q}\cdot C^{-1}\cdot \mathbf{q}=\frac{(K\gamma+k\Gamma)((K^2\alpha^2+({\Gamma\upsilon})^2\beta^2+K\beta(-2{\Gamma\upsilon} \alpha+k\beta))\gamma+k(K\alpha^2+k\beta^2)\Gamma)}{(K^2\gamma^2+(({\Gamma\upsilon})^2+2kK)\gamma\Gamma+k^2\Gamma^2)}
\end{equation}		
\end{widetext}
{{where $C$ is defined as $C=\langle\mathbf{q}(t)\mathbf{q}(t)\rangle$ in the $\lim {t\to \infty}$.}}
\section{Definition of action }
\label{appb}
Given an equation of the form
\begin{align}
\label{defac}
\dot{\mathbf{q}}= \mathbf{D}(\mathbf{q},t)+\mathbf{N}(t)
\end{align}
where $\mathbf{q}$ is the column vector of dynamical variables, $ \mathbf{D}(\mathbf{q},t)$ is the {column vector of systematic forces}, and $\mathbf{N}$ is the column of additive Gaussian noise with zero mean and variance 
\begin{align}
\label{defno}
\langle \mathbf{N}({t}) \mathbf{N}({t'})\rangle=\mathbf{M}\delta(t-t')
\end{align}

Following Refs.\cite{markovich2021thermodynamics,lau2007state,cugliandolo2017rules}, the path probability  $P= e^{-A}$ for  the paths $\mathbf{q}_s$ ($s \in (0,t)$) solving  Eqs.~(\ref{defac}), (\ref{defno}) is governed by the action
\begin{align}
A=\int_0^t\left[
\frac{1}{2}\left(\dot{\mathbf{q}}-\mathbf{D}\right)\mathbf{M}^{-1}\left(\dot{\mathbf{q}}-\mathbf{D}\right)^T+\frac{1}{2}\partial_{\mathbf{q}}\cdot\mathbf{D}\right] ds 
\end{align}
The last term is the Jacobian arising  in the  Stratonovich convention for the discretization, Eq.~(5.15) in Ref.~\cite{lau2007state}, when the weight for the noise history is written in terms of the dynamical variables. Its antisymmetric part under time reversal $\mathcal{T}$ is related to entropy,  its symmetric part to frenesy \cite{maes2020frenesy}.

\section{Entropy production}
\label{appf}
{
From the equations of motion (\ref{modf1mom}),(\ref{modf1aux}) of Sec.~\ref{sec3}
we obtain the action 
\begin{equation}
\label{act1}
\begin{aligned}
A=&\int dt\,\frac{1}{4}\Bigg[\Gamma\left(\dot{\alpha} +\frac{1}{\Gamma}\partial_{\alpha} \Phi -\frac{{\upsilon\Gamma}}{K\gamma+k\Gamma}\partial_\beta{\Phi}\right)^2\\&+\gamma\left(\dot{\beta} +\frac{{\upsilon\Gamma}}{K\gamma+k\Gamma}\partial_{\alpha} \Phi+\frac{1}{\gamma} {\partial_\beta {\Phi}}\right)^2-\frac{2}{\Gamma}\partial_\alpha^2\Phi-\frac{2}{\gamma}\partial_\beta^2\Phi \Bigg]
\end{aligned}
\end{equation}
We can rewrite it as
\begin{equation}
\label{act2}
\begin{split}
& \frac{1}{4} \int dt \Bigg(\Gamma\dot{\alpha}^2+\gamma\dot{\beta}^2+M (\partial_{\alpha} \Phi)^2+N(\partial_{\beta} \Phi)^2\\&+ 2\dot{\alpha}\partial_{\alpha} \Phi+2\dot{\beta}\partial_\beta {\Phi}-\frac{2{\upsilon\Gamma}\Gamma}{K\gamma+k\Gamma}\dot{\alpha}\partial_\beta{\Phi}+\frac{2{\upsilon\Gamma}\gamma}{K\gamma+k\Gamma}\dot{\beta}\partial_{\alpha} \Phi\\&-\frac{2}{\Gamma}\partial_\alpha^2\Phi-\frac{2}{\gamma}\partial_\beta^2\Phi\Bigg)
 \end{split}
\end{equation}
where $M=\frac{1}{\Gamma}+\gamma\left(\frac{{\upsilon\Gamma}}{K\gamma+k\Gamma}\right)^2$ and $N=\frac{1}{\gamma}+\Gamma\left(\frac{{\upsilon\Gamma}}{K\gamma+k\Gamma}\right)^2$, to arrive at}
\begin{equation}
\begin{aligned}
A= &\int dt\frac{1}{4} \Bigg(\Gamma\dot{\alpha}^2+\gamma\dot{\beta}^2+M (\partial_{\alpha} \Phi)^2+N(\partial_{\beta} \Phi)^2+2\frac{d\Phi}{dt}\\&+\frac{2{\upsilon\Gamma}}{K\gamma+k\Gamma}\left(\gamma\dot{\beta}\partial_{\alpha} \Phi-\Gamma\dot{\alpha}\partial_\beta{\Phi}\right)-\frac{2}{\Gamma}\partial_\alpha^2\Phi-\frac{2}{\gamma}\partial_\beta^2\Phi\Bigg)
\end{aligned}
\end{equation}

{If both noise have equal strength, i.e. $\Gamma=\gamma$,
\begin{equation}
\begin{aligned}
A= &\int dt\frac{1}{4} \Bigg(\gamma(\dot{\alpha}^2+\dot{\beta}^2)+M( (\partial_{\alpha} \Phi)^2+(\partial_{\beta} \Phi)^2)+2\frac{d\Phi}{dt}\\&+\frac{2{\upsilon}\gamma}{K+k}\left(\dot{\beta}\partial_{\alpha} \Phi-\dot{\alpha}\partial_\beta{\Phi}\right)-\frac{2}{\gamma}\left(\partial_\alpha^2\Phi+\partial_\beta^2\Phi\right)\Bigg)
\end{aligned}
\end{equation}
where the second last parenthesis can also be written as $\partial_{\bm{\mathcal{C}}} \Phi\wedge \dot{\bm{\mathcal{C}}}$.
{Since $\alpha$ and $\beta$ have opposite time reversal signature}, the time antisymmetric part of the action,$A-A^R$, reads
\begin{equation}
\begin{split}
\int dt\frac{1}{4} \Bigg(M( (\partial_{\alpha} \Phi)^2+(\partial_{\beta} \Phi)^2-(\partial_{\alpha} {\Phi}^R)^2-(\partial_{\beta} {\Phi}^R)^2)\\+2\frac{d(\Phi+{\Phi}^R)}{dt}-\frac{2}{\gamma}\left(\partial_\alpha^2(\Phi-{\Phi}^R)+\partial_\beta^2(\Phi-{\Phi}^R)\right)\\+\frac{2{\upsilon}\gamma}{K+k}\left(\dot{\beta}\partial_{\alpha} (\Phi-{\Phi}^R)-\dot{\alpha}\partial_\beta({\Phi}-{\Phi}^R)\right)\Bigg)
\end{split}
\end{equation}
where ${\Phi}^R\equiv\mathcal{T}\Phi$ and $A^R\equiv\mathcal{T}A$. With $\Phi_s=(\Phi+{\Phi}^R)/2$ and $\Phi_a=(\Phi-{\Phi}^R)/2$ for the $\mathcal{T}$-even and $\mathcal{T}$-odd parts of the NH, we have
\begin{equation}
\begin{aligned}
A-A^R &= \int dt \Bigg(M\partial_{\bm{\mathcal{C}}}\Phi_a\cdot \partial_{\bm{\mathcal{C}}}\Phi_s+\frac{d\Phi_s}{dt}\\&+\frac{{\upsilon}\gamma}{K+k}\left(\dot{\beta}\partial_{\alpha} \Phi_a-\dot{\alpha}\partial_\beta{\Phi}_a\right)-\frac{1}{\gamma}\partial_{\bm{\mathcal{C}}}^2\Phi_a\Bigg)
\end{aligned}
\end{equation} 
or, equivalently,
\begin{equation}
\label{entfin}
\int dt \Bigg(M\partial_{\bm{\mathcal{C}}}\Phi_a\cdot \partial_{\bm{\mathcal{C}}}\Phi_s+\frac{d\Phi_s}{dt}+\frac{{\upsilon}\gamma}{K+k}\partial_{\bm{\mathcal{C}}} \Phi_a\wedge\dot{\bm{\mathcal{C}}}-\frac{1}{\gamma}\partial_{\bm{\mathcal{C}}}^2\Phi_a\Bigg)
\end{equation}
}

{The generic case, when the noises have different strength, can also be written in the form above by rescaling $\alpha\rightarrow {\alpha}/{\sqrt{\Gamma}}$ and $\beta\rightarrow {\beta}/{\sqrt{\gamma}}$, upon which Eq.(\ref{act2}) becomes
\begin{equation}
\begin{aligned}
A= &\int dt\frac{1}{4} \Bigg(\dot{\alpha}^2+\dot{\beta}^2+\tilde{M}( (\partial_{\alpha} \Phi')^2+(\partial_{\beta} \Phi')^2)+2\frac{d\Phi'}{dt}\\&+\frac{2{\upsilon\Gamma}\sqrt{\Gamma\gamma}}{K\gamma+k\Gamma}\left(\dot{\beta}\partial_{\alpha} \Phi'-\dot{\alpha}\partial_\beta{\Phi'}\right)-2(\partial_\alpha^2\Phi+\partial_\beta^2\Phi)\Bigg)
\end{aligned}
\end{equation}
where $\Phi'=\Phi({\alpha}/{\sqrt{\Gamma}},{\beta}/{\sqrt{\gamma}})$ and $\tilde{M}=1+\Gamma\gamma\left(\frac{\upsilon}{K\gamma+k\Gamma}\right)^2$. Following the arguments above, the antisymetric part $A-A^R$ of the action is
\begin{equation}
 \!\int \! dt \Bigg(\tilde{M}\partial_{\bm{\mathcal{C}}}\Phi'_a\cdot \partial_{\bm{\mathcal{C}}}\Phi'_s+\frac{d\Phi'_s}{dt}+\frac{{\upsilon\Gamma}\sqrt{\Gamma\gamma}}{K\gamma+k\Gamma}\partial_{\bm{\mathcal{C}}} \Phi'_a\wedge\dot{\bm{\mathcal{C}}}-\partial_{\bm{\mathcal{C}}}^2\Phi'_a\Bigg)
\end{equation}
}

{Now we explicitly calculate the terms in (\ref{entfin}) for the NH given in (\ref{pot1}):
\begin{equation}
\Phi_a=\frac{-(K\gamma+k\Gamma)(K\gamma{\upsilon\Gamma}\beta \alpha)}{(K^2\gamma^2+(({\upsilon\Gamma})^2+2kK)\gamma\Gamma+k^2\Gamma^2)}
\end{equation}
\begin{widetext}
\begin{equation}
\partial_{\bm{\mathcal{C}}} \Phi_a=\frac{-(K\gamma+k\Gamma)K\gamma{\upsilon\Gamma}}{(K^2\gamma^2+(({\upsilon\Gamma})^2+2kK)\gamma\Gamma+k^2\Gamma^2)}\left(\beta,\alpha\right)\equiv \frac{N''}{2}\left(\beta,\alpha\right);\quad\partial^2_{\bm{\mathcal{C}}} \Phi_a=0
\end{equation}
\begin{equation}
\Phi_s=\frac{(K\gamma+k\Gamma)((K^2\alpha^2+({\upsilon\Gamma})^2\beta^2+Kk\beta^2)\gamma+k(K\alpha^2+k\beta^2)\Gamma)}{2(K^2\gamma^2+(({\upsilon\Gamma})^2+2kK)\gamma\Gamma+k^2\Gamma^2)}
\end{equation}
\begin{equation}
\partial_{\bm{\mathcal{C}}} \Phi_s=\frac{(K\gamma+k\Gamma)}{(K^2\gamma^2+(({\upsilon\Gamma})^2+2kK)\gamma\Gamma+k^2\Gamma^2)}\left(K^2\gamma\alpha+kK\Gamma\alpha,({\upsilon\Gamma})^2\gamma\beta+kK\gamma\beta+k^2\Gamma\beta\right)
\end{equation}
\begin{equation}
\left(\partial_{\alpha} \Phi_s\partial_{\alpha} \Phi_a, \partial_{\beta} \Phi_s\partial_{\beta} \Phi_a\right)=\frac{-(K\gamma+k\Gamma)^2K\gamma{\upsilon\Gamma}}{(K^2\gamma^2+(({\upsilon\Gamma})^2+2kK)\gamma\Gamma+k^2\Gamma^2)^2}\left((K^2\gamma+kK\Gamma)\alpha\beta,  (({\upsilon\Gamma})^2\gamma+kK\gamma+k^2\Gamma)\alpha\beta\right)
\end{equation}
\begin{equation}
\begin{split}
\partial_{\bm{\mathcal{C}}}\Phi_s\cdot\partial_{\bm{\mathcal{C}}}\Phi_a=\frac{-(K\gamma+k\Gamma)^2K\gamma{\upsilon\Gamma}}{(K^2\gamma^2+(({\upsilon\Gamma})^2+2kK)\gamma\Gamma+k^2\Gamma^2)^2}\left((K^2\gamma+kK\Gamma)+ (({\upsilon\Gamma})^2\gamma+kK\gamma+k^2\Gamma)\right)\alpha\beta \equiv \frac{M'}{M}\alpha\beta
\end{split}
\end{equation}
\begin{equation}
\begin{split}
\dot{\beta}\partial_{\alpha} \Phi_a-\dot{\alpha}\partial_{\beta} \Phi_a=
-N'\frac{d\beta^2}{dt}+N''\frac{d\alpha^2}{dt}= N''\frac{d(\alpha^2-\beta^2)}{dt};\quad N'\equiv N''\frac{{\upsilon\Gamma}\gamma}{K\gamma+k\Gamma}
\end{split}
\end{equation}
\end{widetext}
}

\section{Generic case}
\label{appg}
{
Here we display the structure of entropy production for a generic equilibrium-like Langevin equations. We recall the generic equilibrium-like Langevin equations (\ref{noi1}), (\ref{lang1}), and restrict to additive noise for simplicity.
\begin{align}
\dot{\bm{\mathcal{C}}}=-(\bm{Q}+\bm{L^a})\cdot\partial_{\bm{\mathcal{C}}}\phi+\partial_{\bm{\mathcal{C}}}\bm{L^a}+\bm{g}^i\xi_i
\end{align}
\begin{align}
\langle \xi_i(t)\xi_j(t')\rangle=2\delta_{ij}\delta(t-t') \quad \bm{Q}=\bm{g}^i\bm{g}^i
\end{align}
For simplicity, we restrict the discussion $\partial_{\bm{\mathcal{C}}}\bm{L^a}=0$, as often the case \cite{chaikin1995principles} and $\partial_{\bm{\mathcal{C}}}\cdot(\bm{L^a}\cdot\partial_{\bm{\mathcal{C}}}\phi)=0$, both are always satisfied for constant $\bm{L^a}$. Then, the action $A$ is
\begin{equation}
\begin{aligned}
\int \! dt  \left[\dot{\bm{\mathcal{C}}}+(\bm{Q}+\bm{L^a})\cdot{\partial_{\bm{\mathcal{C}}} \phi}\right]^T 
\!\! \cdot {\bm{Q}}^{-1} \cdot\left[\dot{\bm{\mathcal{C}}}+(\bm{Q}+\bm{L^a})\cdot{\partial_{\bm{\mathcal{C}}} \phi}\right]\\-\bm{Q}:\partial_{\bm{\mathcal{C}}} \partial_{\bm{\mathcal{C}}}\phi
\end{aligned}
\end{equation} 
Using $(\bm{AB})^T=\bm{B^TA^T}$ and the fact that $\bm{Q}$ and $\bm{Q}^{-1}$ are symmetric matrices and $\left[{\partial_{\bm{\mathcal{C}}} \phi}\right]^T\cdot\bm{\bm{L^a}}\cdot{\partial_{\bm{\mathcal{C}}} \phi}=0$, because $\bm{L^a}$ is antisymmetric, we can write 
\begin{equation}
\begin{aligned}
A & =\int dt  \Bigg(\dot{\bm{\mathcal{C}}}^T\cdot\bm{\bm{Q}}^{-1}\cdot\dot{\bm{\mathcal{C}}}+\dot{\bm{\mathcal{C}}}^T\cdot\bm{\bm{Q}}^{-1}\cdot\bm{\bm{Q}}\cdot{\partial_{\bm{\mathcal{C}}} \phi}\\&+\dot{\bm{\mathcal{C}}}^T\cdot\bm{\bm{Q}}^{-1}\cdot\bm{\bm{L^a}}\cdot{\partial_{\bm{\mathcal{C}}} \phi}+\left[\bm{\bm{Q}}\cdot{\partial_{\bm{\mathcal{C}}} \phi}\right]^T\cdot \bm{\bm{Q}}^{-1}\cdot \dot{\bm{\mathcal{C}}}\\&+\left[\bm{\bm{Q}}\!\!\cdot{\partial_{\bm{\mathcal{C}}} \phi}\right]^T\!\!\cdot\bm{\bm{Q}}^{-1}\!\!\cdot\bm{\bm{Q}}\!\!\cdot{\partial_{\bm{\mathcal{C}}} \phi}+\left[\bm{\bm{Q}}\cdot{\partial_{\bm{\mathcal{C}}} \phi}\right]^T\cdot\bm{\bm{Q}}^{-1}\!\!\cdot\bm{\bm{L^a}}\cdot{\partial_{\bm{\mathcal{C}}} \phi} \\&
+\left[\bm{\bm{L^a}}\cdot{\partial_{\bm{\mathcal{C}}} \phi}\right]^T\!\!\cdot\bm{\bm{Q}}^{-1}\cdot\bm{\bm{L^a}}\cdot{\partial_{\bm{\mathcal{C}}} \phi}+\left[\bm{\bm{L^a}}\cdot{\nabla_{\bm{\mathcal{C}}} \phi}\right]^T\!\!\cdot\bm{\bm{Q}}^{-1}\cdot\dot{\bm{\mathcal{C}}}
\\ &+\left[\bm{\bm{L^a}}\cdot{\partial_{\bm{\mathcal{C}}} \phi}\right]^T\!\!\cdot\bm{\bm{Q}}^{-1}\cdot\bm{\bm{Q}}\!\!\cdot{\partial_{\bm{\mathcal{C}}} \phi}-\bm{Q}:\partial_{\bm{\mathcal{C}}} \partial_{\bm{\mathcal{C}}}\phi\Bigg) \\
%&=\int dt \Bigg(\left[\dot{\bm{\mathcal{C}}}\right]^T\bm{\bm{Q}}^{-1}\dot{\bm{\mathcal{C}}}+\left[\dot{\bm{\mathcal{C}}}\right]^T\cdot{\partial_{\bm{\mathcal{C}}} \phi}+\left[\partial_{\bm{\mathcal{C}}} \phi\right]^T\cdot\dot{\bm{\mathcal{C}}}\\&+\left[{\partial_{\bm{\mathcal{C}}} \phi}\right]^T\cdot\bm{\bm{Q}}\cdot{\partial_{\bm{\mathcal{C}}} \phi}-\left[{\partial_{\bm{\mathcal{C}}} \phi}\right]^T\cdot\bm{\bm{L^a}}\bm{\bm{Q}}^{-1}\dot{\bm{\mathcal{C}}}\\&+\left[\dot{\bm{\mathcal{C}}}\right]^T\bm{\bm{Q}}^{-1}\bm{\bm{L^a}}\cdot{\partial_{\bm{\mathcal{C}}} \phi}-\left[{\partial_{\bm{\mathcal{C}}} \phi}\right]^T\cdot\bm{\bm{L^a}}\bm{\bm{Q}}^{-1}\bm{\bm{L^a}}\cdot{\partial_{\bm{\mathcal{C}}} \phi}\Bigg) \\
&=\int dt \Bigg(\dot{\bm{\mathcal{C}}}^T\cdot\bm{\bm{Q}}^{-1}\cdot\dot{\bm{\mathcal{C}}}+2\dot{\phi}+\left[{\partial_{\bm{\mathcal{C}}} \phi}\right]^T\!\cdot\bm{\bm{Q}}\cdot{\partial_{\bm{\mathcal{C}}} \phi}\\&-\left[{\partial_{\bm{\mathcal{C}}} \phi}\right]^T\!\cdot\bm{\bm{L^a}}\cdot\bm{\bm{Q}}^{-1}\cdot\dot{\bm{\mathcal{C}}} +\dot{\bm{\mathcal{C}}}^T\cdot\bm{\bm{Q}}^{-1}\cdot\bm{\bm{L^a}}\cdot{\partial_{\bm{\mathcal{C}}} \phi}\\&-\left[{\partial_{\bm{\mathcal{C}}} \phi}\right]^T\!\cdot\bm{\bm{L^a}}\cdot\bm{\bm{Q}}^{-1}\cdot\bm{\bm{L^a}}\cdot{\partial_{\bm{\mathcal{C}}} \phi}-\bm{Q}:\partial_{\bm{\mathcal{C}}} \partial_{\bm{\mathcal{C}}}\phi\Bigg)
\end{aligned}
\end{equation}
}

{When the noises have equal strength but no cross correlations, i.e.,
$Q_{ij}\propto\delta_{ij}$. then
\begin{equation}
\label{actf}
%\begin{aligned}
A\propto \int \!dt \Big[\dot{\bm{\mathcal{C}}}^2+2\dot{\phi}+(\partial_{\bm{\mathcal{C}}} \phi)^2 +2\dot{\bm{\mathcal{C}}}\cdot\bm{\bm{L^a}}\cdot{\partial_{\bm{\mathcal{C}}} \phi}+(\bm{\bm{L^a}}\cdot{\partial_{\bm{\mathcal{C}}} \phi})^2- \partial_{\bm{\mathcal{C}}}^2\phi\Big]
%\end{aligned}
\end{equation}
}
{As discussed above, the general case of different noise strengths reduces to this form upon variable rescaling.}

{To proceed further we choose for $\bm{L^a}$ a matrix, $\bm{J}$ (or $\tilde{\bm{J}}$ if singular and odd dimensional ), which can generate any antisymmetric matrix through the similarity transformation (see App.~\ref{apph}),
where $\bm{J}$ is a $2n\times 2n$ matrix written in $2 \times 2$ block form  \cite{thompson1988normal,eves1980elementary,schwerdtfeger1961introduction}:
\begin{equation}
\label{assym}
\bm{J}\equiv \text{diag}\Bigg \{\begin{pmatrix}
0 & 1 \\
-1 & 0
\end{pmatrix},\, \begin{pmatrix}
0 & 1 \\
-1 & 0
\end{pmatrix},...,\begin{pmatrix}
0 & 1 \\
-1 & 0
\end{pmatrix}\Bigg\}
\end{equation}
Taking $\bm{L^a}=\bm{J}$ gives $(\bm{\bm{L^a}}\cdot{\partial_{\bm{\mathcal{C}}} \phi})^2=({\partial_{\bm{\mathcal{C}}} \phi})^2$. As a result (\ref{actf}) reduces to
\begin{equation}
\label{eqf7}
%\begin{aligned}
A\propto \int \!dt \Big[\dot{\bm{\mathcal{C}}}^2+2\dot{\phi}+2(\partial_{\bm{\mathcal{C}}} \phi)^2 +2\sum\limits_{block}({\partial_{\bm{\mathcal{C}}} \phi}\wedge\dot{\bm{\mathcal{C}}})_{ij}- \partial_{\bm{\mathcal{C}}}^2\phi\Big]
%\end{aligned}
\end{equation}
where $\dot{\mathcal{C}}_i\partial_{\mathcal{C}_j}\phi-\dot{\mathcal{C}}_j\partial_{\mathcal{C}_i}\phi\equiv({\partial_{\bm{\mathcal{C}}} \phi}\wedge\dot{\bm{\mathcal{C}}})_{ij}$  giving $\dot{\bm{\mathcal{C}}}\cdot\bm{\bm{J}}\cdot{\partial_{\bm{\mathcal{C}}} \phi}=\sum\limits_{block}({\partial_{\bm{\mathcal{C}}} \phi}\wedge\dot{\bm{\mathcal{C}}})_{ij}$, and $A-A^R$ takes the form
\begin{equation}
\label{genent}
\begin{aligned}
\propto\int dt\Big[\dot{\phi}_s+2\partial_{\bm{\mathcal{C}}} \phi_s\cdot\partial_{\bm{\mathcal{C}}} \phi_a+\sum\limits_{blocks}({\partial_{\bm{\mathcal{C}}} \phi_l}\wedge\dot{\bm{\mathcal{C}}})_{ij}-2\partial_{\bm{\mathcal{C}}}^2\phi_a\Big]
\end{aligned}
\end{equation}
Here, $l$ stands for either $a$ or $s$, depending on whether $\mathcal{C}_i$ and $\mathcal{C}_j$  have opposite or same signatures under $\mathcal{T}$.
}

\section{Properties of antisymmetric matrices}
\label{apph}
{If $\bm{M}$ is an even-dimensional non-singular $2n \times 2n$ antisymmetric matrix, then there exists a non-singular $2n \times 2n$ matrix $\bm{P}$ such that
\begin{equation}
\label{con}
\bm{M} = \bm{P^T\cdot J\cdot P }
\end{equation}
where the $2n \times 2n$ matrix $\bm{J}$ written in $2 \times 2$ block form is given by Eq.~(\ref{assym})
}

{If $\bm{M}$ is a singular antisymmetric $d \times d$ matrix of rank $2n$ (where $d$ is either even or odd and $d > 2n$), then there exists a non-singular $d \times d$ matrix $P$ such that
\begin{equation}
\label{con1}
\bm{M = P^{T}\cdot \tilde{J}\cdot P }
\end{equation}
and $\bm{\tilde{J}}$ is the $d \times d$ matrix that is given in block form by
\[\bm{\tilde{J}}\equiv
\begin{pmatrix}
  \begin{matrix}
  \bigj
  \end{matrix}
  & \rvline & \bigzero \\
\hline
  \bigzero & \rvline &
  \begin{matrix}
  \bigzero
  \end{matrix}
\end{pmatrix}
\]
where the $2n\times 2n$ matrix $\bm{J}$ is defined in Eqn.~(\ref{assym}) and $\bm{0}$ is a zero matrix of the appropriate number of rows and columns. Note that if $d = 2n$, then Eq.~ (\ref{con1}) reduces to Eq.~(\ref{con}).
}

{Two matrices , $\bm{M}$ and $\bm{J}$ (or $\bm{\tilde{J}}$) are said to be congruent (similar) if they are related by Eqn.~(\ref{con}) ( \ref{con1}). Thus Eq.~(\ref{con}) (\ref{con1}) imply that all $d \times d$ antisymmetric matrices of rank $2n$ (where $n\le d/2 $) belong to the same congruent class, which is uniquely specified by $d$ and $n$. Thus all the antisymmetric matrices of a given $d$ and $n$ can be generated from their respective $\bm{J}$ ($\bm{\tilde{J}}$) through a similarity transformation.}

\section{Noise as a source of entropy production}
\label{appl}
{We recall the generic equilibrium-like Langevin equations (\ref{noi1}), (\ref{lang1}) and  restrict to $\partial_{\bm{\mathcal{C}}}\bm{L^a}=0$ as above.
\begin{align}
\dot{\mathcal{C}}_\mu=-(Q_{\mu\nu}+L^a_{\mu\nu})\partial_{\bm{\mathcal{C}}_{{\nu}}}\phi+g_\mu^i(\bm{\mathcal{{{C}}}})\xi_i
\end{align}
Time derivative of the NH is
\begin{align}
\dot{\phi}=\dot{\mathcal{C}}_\mu\partial_{\bm{\mathcal{C}}_{{\mu}}}\phi
\end{align}
There is no systematic change in the value of the NH  due to the  ``reactive'' part of the dynamics, as $\dot{\phi}=-L^a_{\mu\nu}\partial_{\bm{\mathcal{C}}_{{\nu}}}\phi \partial_{\bm{\mathcal{C}}_{{\mu}}}\phi=0$, because $L^a_{\mu\nu}$ is antisymmetric. Therefore, dynamics due to ``reactive'' part is along the level surfaces of NH. 
The temporal change in $\phi$ due to ``dissipative" part is 
\begin{align}
\dot{\phi}=-Q_{\mu\nu}\partial_{\bm{\mathcal{C}}_{{\nu}}}\phi \partial_{\bm{\mathcal{C}}_{{\mu}}}\phi<0
\end{align}
Because noise covariance matrix, $Q_{\mu\nu}$, is symmetric and positive definite (semi-definite in general), the ``dissipative" term will always systematically decrease the NH.  Since the NH is bounded from below the systematic dynamics will lead to the minimum of  the NH, in accord with the notion of a Lyapunov function. At the minimum the derivative of the NH vanishes, which implies the vanishing of all ``dissipative" and ``reactive'' forces. But fluctuation keep kicking the system out of the minimum which leads to a finite ``reactive" (and ``dissipative" ) force resulting in the movement along the contour lines and producing entropy. It is important to note that it is the noise which is responsible for the entropy production (especially when both $\alpha$ and $\beta$ are even under $\mathcal{T}$). This scenario is totally different from equilibrium where fluctuations due to noise do not give rise to any dissipation. Therefore, the above ``reactive" entropy producing fluctuations are suitably addressed as ``dissipative fluctuations".}

\section{Calulation for excess frenesy}
\label{appc}
{ The equations of motion (\ref{modf1mom}) and (\ref{modf1aux}) in the presence of perturbing force, i.e., taking $\phi\rightarrow \phi-f_iq_i$ with $q_1=\alpha$ and $q_2=\beta$, given by
\begin{equation}
\dot{\alpha} =-\frac{1}{\Gamma}\partial_{\alpha} \Phi +\frac{{\upsilon\Gamma}}{K\gamma+k\Gamma}\partial_\beta{\Phi}+\frac{f_1}{\Gamma}-\frac{{\upsilon\Gamma} f_2}{K\gamma+k\Gamma}+\xi_\alpha
\end{equation}
\begin{equation}
\dot{\beta} =-\frac{{\upsilon\Gamma}}{K\gamma+k\Gamma}\partial_{\alpha} \Phi-\frac{1}{\gamma} {\partial_\beta {\Phi}}+\frac{{\upsilon\Gamma} f_1}{K\gamma+k\Gamma}+\frac{f_2}{\gamma}+{\xi}_\beta
\end{equation}
We define the perturbation part $A_f$ of the action by
\begin{equation}
A=A_0+A_f
\end{equation}
where $A$ is the total action in presence of the external forces $f_i$ and $A_0$ is the action in the absence.
As shown in App.~\ref{appb}, the path probability for the trajectory $\omega=(X_s)$ $(s \in [0,t])$ is given by
\begin{equation}
P[\omega]=e^{-A}=e^{-A_f(\omega)}P_{ref}[\omega]
\end{equation}
where $P_\text{ref}[\omega]=e^{-A_0(\omega)}$ is the path probability in the absence of the perturbation. They are given by 
\begin{equation}
\begin{aligned}
A_0=&\int ds\,\frac{1}{4}\Bigg[\Gamma\left(\dot{\alpha} +\frac{1}{\Gamma}\partial_{\alpha} \Phi -\frac{{\upsilon\Gamma}}{K\gamma+k\Gamma}\partial_\beta{\Phi}\right)^2\\&+\gamma\left(\dot{\beta} +\frac{{\upsilon\Gamma}}{K\gamma+k\Gamma}\partial_{\alpha} \Phi+\frac{1}{\gamma} {\partial_\beta {\Phi}}\right)^2-\frac{2}{\Gamma}\partial_\alpha^2\Phi-\frac{2}{\gamma}\partial_\beta^2\Phi\Bigg]
\end{aligned}
\end{equation}
and
\begin{equation}
\begin{aligned}
&A_f=\\&\int ds\,{ \Gamma\gamma}\Bigg[\frac{1}{\gamma}\left(\!\dot{\alpha}\! +\!\frac{1}{\Gamma}\partial_{\alpha} \Phi \!-\!\frac{{\upsilon\Gamma}}{K\gamma+k\Gamma}\partial_\beta{\Phi}\!\right)\left(\frac{f_1}{\Gamma}\!-\!\frac{{\upsilon\Gamma} f_2}{K\gamma+k\Gamma}\right)\\&+\frac{1}{\Gamma}\left(\dot{\beta} +\frac{{\upsilon\Gamma}}{K\gamma+k\Gamma}\partial_{\alpha} \Phi+\frac{1}{\gamma} {\partial_\beta {\Phi}}\right)\left(\frac{{\upsilon\Gamma} f_1}{K\gamma+k\Gamma}+\frac{f_2}{\gamma}\right)\Bigg]
\end{aligned}
\end{equation}
which can be rewritten as
\begin{equation}
\begin{aligned}
A_f=&\int dt\,\Bigg[\left(\frac{1}{\Gamma}+\frac{({\upsilon\Gamma})^2\gamma }{(K\gamma+k\Gamma)^2}\right)\left(\frac{f_1}{\Gamma}\partial_{\alpha}\Phi+\frac{f_2}{\gamma}\partial_{\beta}\Phi\right)\\&+\Gamma\dot{\alpha}\left(\frac{f_1}{\Gamma}-\frac{{\upsilon\Gamma} f_2}{K\gamma+k\Gamma}\right)+\gamma\dot{\beta}\left(\frac{{\upsilon\Gamma} f_1}{K\gamma+k\Gamma}+\frac{f_2}{\gamma}\right)\Bigg]
\end{aligned}
\end{equation}
When $\alpha$ and $\beta$ are $\mathcal{T}$-even
\begin{equation}
\begin{aligned}
&A_f+\mathcal{T}A_f=\\&\int dt\,2\left(\frac{1}{\Gamma}+\frac{({\upsilon\Gamma})^2\gamma }{(K\gamma+k\Gamma)^2}\right)\left(\frac{f_1}{\Gamma}\partial_{\alpha}\Phi+\frac{f_2}{\gamma}\partial_{\beta}\Phi\right)
\end{aligned}
\end{equation}
After scaling, $\alpha\rightarrow {\alpha}/{\sqrt{\Gamma}}$ and $\beta\rightarrow {\beta}/{\sqrt{\gamma}}$,$\Phi'\equiv \Phi({\alpha}/{\sqrt{\Gamma}},{\beta}/{\sqrt{\gamma}})$, we have
\begin{equation}
\begin{aligned}
A_f+\mathcal{T}A_f=&\int ds\,2\Bigg[\left(\frac{1}{\Gamma}+\frac{({\upsilon\Gamma})^2\gamma }{(K\gamma+k\Gamma)^2}\right)\mathbf{f}\cdot\partial_{\bm{\mathcal{C}}}\Phi'\Bigg]
\end{aligned}
\end{equation}
The antisymmetric part of $A_f$
\begin{equation}
\begin{aligned}
&A_f-\mathcal{T}A_f=\\&\int ds\,2\Bigg[\Gamma\dot{\alpha}\left(\frac{f_1}{\Gamma}-\frac{{\upsilon\Gamma} f_2}{K\gamma+k\Gamma}\right)+\gamma\dot{\beta}\left(\frac{{\upsilon\Gamma} f_1}{K\gamma+k\Gamma}+\frac{f_2}{\gamma}\right)\Bigg]
\end{aligned}
\end{equation}
When $\alpha$ is $\mathcal{T}$-even and $\beta$ is $\mathcal{T}$-odd, we find
\begin{equation}
\begin{aligned}
&A_f+\mathcal{T}A_f=\\&\int dt\,2\Bigg[\left(\frac{1}{\Gamma}+\frac{({\upsilon\Gamma})^2\gamma }{(K\gamma+k\Gamma)^2}\right)\left(\frac{f_1}{\Gamma}\partial_{\alpha}\Phi_s+\frac{f_2}{\gamma}\partial_{\beta}\Phi_a\right)\\&+\gamma\dot{\beta}\left(\frac{{\upsilon\Gamma} f_1}{K\gamma+k\Gamma}+\frac{f_2}{\gamma}\right)\Bigg]
\end{aligned}
\end{equation}
or, more explicitaly,
\begin{equation}
\begin{aligned}
&A_f+\mathcal{T}A_f =\int_0^t\gamma\Gamma \left[\frac{1}{\gamma}\left(\frac{f_1}{\Gamma}-\frac{{\upsilon\Gamma}}{K\gamma+\Gamma}f_2\right)\left(\frac{K\alpha}{\Gamma}\right)\right]  \\&  +\left[\frac{1}{\Gamma}\left(\frac{{\upsilon\Gamma}}{K\gamma+\Gamma} f_1+\frac{f_2}{\gamma}\right)\dot{\beta}\right]ds
\end{aligned}
\end{equation}
and,
\begin{equation}
\begin{aligned}
&A_f-\mathcal{T}A_f=\\&\int ds\,2\Bigg[\left(\frac{1}{\Gamma}+\frac{({\upsilon\Gamma})^2\gamma }{(K\gamma+k\Gamma)^2}\right)\left(\frac{f_1}{\Gamma}\partial_{\alpha}\Phi_a+\frac{f_2}{\gamma}\partial_{\beta}\Phi_s\right)\\&+\Gamma\dot{\alpha}\left(\frac{f_1}{\Gamma}-\frac{{\upsilon\Gamma} f_2}{K\gamma+k\Gamma}\right)\Bigg]
\end{aligned}
\end{equation}
which for our $\Phi$, reduces to
\begin{equation}
\begin{aligned}
A_f-\mathcal{T}A_f&=2\int_0^t\gamma\Gamma \left[\frac{1}{\gamma}\left(\frac{f_1}{\Gamma}-\frac{\upsilon\Gamma}{K\gamma+\Gamma}f_2\right)\left(\dot{\alpha} -\frac{\upsilon\Gamma \beta}{\Gamma}\right)\right]  \\&  +\left[\frac{1}{\Gamma}\left(\frac{{\upsilon\Gamma}}{K\gamma+\Gamma} f_1+\frac{f_2}{\gamma}\right) \frac{k\beta}{\gamma}\right]ds
\end{aligned}
\end{equation}
}

\section{Directly perturbing the EOM vs perturbing the NH}
\label{appd}
The usual way to perturb nonequilibrium systems (without a Hamiltonian) is by introducing a perturbing force $f$ into the equations of motion for  $X$~\cite{maes2020response,dadhichi2018origins}
\begin{equation}
\label{appmod2mom}
\Gamma \dot{X} =-{\partial_X H}+f +{ x}
\end{equation}
\begin{equation}
\label{appmod2aux}
\gamma \dot{x} =- x+{\xi}_x
\end{equation}
Quite different results are obtained if one exploits Graham's equilibrium-like structure and perturbs the dynamic variable $X$ by adding a term $\phi\rightarrow \phi-f_1X$ to the NH. For a harmonic NH, the latter strategy results in
\begin{equation}
\label{appmod2mom}
\dot{X} =-\frac{KX}{\Gamma} +\frac{x}{\Gamma}
\end{equation}
\begin{equation}
\label{appmod2aux}
\dot{x} =- \frac{x}{\gamma} +\frac{f_1}{K\gamma+\Gamma}+{\xi}_x
\end{equation}

\end{appendix}

%\end{equation}
%\bibliographystyle{ieeetr}
%\bibliography{p_bracket}
\bibliographystyle{apsrev4-1}
\bibliography{main}

\begin{thebibliography}{55}
\bibitem [{b()}]{b}%
  \BibitemOpen
  \href@noop {} {}\bibinfo {note} {{Note that the constraints put on the
  antisymmetric coupling matrix to derive Eq.~(\ref{genent}), in the appendix,
  are always satisfied for a system having only two degrees of
  freedom.}}\BibitemShut {Stop}%
\end{thebibliography}
\end{document}